# Natural CMT2 variation is associated with genome-wide methylation changes and temperature adaptation

Xia Shen[1], Simon Forsberg[1], Mats Pettersson[1], Zheya Sheng[1] and Örjan Carlborg[1]*

Affiliations: [1]Swedish University of Agricultural Sciences, Department of Clinical Sciences, Division of Computational Genetics, Box 7078, SE-750 07 Uppsala, Sweden.

*To whom correspondence should be addressed. Email: orjan.carlborg@slu.se

A central problem when studying adaptation to a new environment is the interplay between genetic variation and phenotypic plasticity [1-3]. *Arabidopsis thaliana* has colonized a wide range of habitats across the world and it is therefore an attractive model for studying the genetic mechanisms underlying environmental adaptation [4,5]. Here, we used publicly available data from two collections of *A. thaliana* accessions, covering the native range of the species[4,6-11], to identify loci associated with differences in climates at the sampling sites. To address the confounding between geographic location, climate and population structure, a new genome-wide association analysis method was developed that facilitates detection of potentially adaptive loci where the alternative alleles display different tolerable climate ranges. Sixteen novel such loci, many of which contained candidate genes with amino acid changes, were found including a strong association between Chromomethylase 2 (*CMT2*) and variability in seasonal temperatures. The reference allele dominated in areas with less seasonal variability in temperature, and the alternative allele, which disrupts genome-wide CHH-methylation, existed in both stable and variable regions. Our results link natural variation in *CMT2*, and differential genome-wide CHH methylation, to the distribution of *A. thaliana* accessions across habitats with different seasonal temperature variability. They also suggest a role for genetic regulation of epigenetic modifications in natural adaptation, potentially through differential allelic plasticity, and illustrate the importance of re-analyses of existing data using new analytical methods to obtain a more complete understanding of the mechanisms contributing to adaptation.

Several large collections of *A. thaliana* accessions have either been whole-genome re-sequenced or high-density SNP genotyped [4,6-11]. The included accessions have adapted to a wide range of different climatic conditions, and therefore loci involved in climate adaptation will display genotype by climate-at-sampling-site correlations in these populations. Genome-wide association analyses will therefore detect signals of natural selection involved in environmental adaptation by screening for associations between marker genotypes and climate at the sampling locations. However, the strong confounding between genetic relationships due to adaptation (selection) and population structure (drift) will result in many false positive signals unless properly accounted for. General kinship corrections account for population structure, but also decrease power to infer adaptive loci for climate conditions that correlate with geographic locations which confound with the general population structure, e.g. those following north-south or east-west geographic clines such as day light and temperature. Such lack of power is inherent to all analyses, regardless of whether associations are sought to the climate-at-sampling-sites or to phenotypes underlying such adaptation that are measured under experimental conditions. Here, we extend and utilize a novel approach [12,13] that instead of mapping loci by differences in allele-frequencies between environments, map adaptive loci based on their distributions across climates using a heterogeneity-of-variance test. Conceptually, the method identifies loci where alleles have different tolerable ranges of climate conditions, and its power is less affected by kinship corrections for population stratification, since such allelic plasticity is not directly confounded with geographical locations and therefore less confounded with population structure.. A genome-wide association analysis was performed for thirteen climate variables across ~215,000 SNPs in 948 *A. thaliana* accessions representing the native range of the species [14]. In total, sixteen genome-wide significant loci were found to be associated with eight climate variables (Table 1), none of which could be found using standard methods for GWAS analyses [4,5,15-17]. The effects were in general quite large, from 0.3 to 0.5 residual standard deviations (Table 1). The detailed results for each trait are reported in Supplementary Figure 1-13.

Utilizing data from the 1001-genomes project [6-11] (http://1001genomes.org), we identified five functional candidate genes (Table 1) and 11 less well characterized genes (Supplementary Table 1) with either missense, nonsense or frameshift mutations in high linkage disequilibrium (LD; $r^2 > 0.8$) with the leading SNPs. 76 additional linked loci or genes without candidate mutations in the coding regions are reported in Supplementary Table 2.

A strong association to temperature seasonality was identified near Chromomethylase 2 (*CMT2*; Table 1, Fig. 1). Temperature seasonality is a measure of the variability in temperature over the year, where stable areas are generally found near large bodies of water (e.g. London near the Atlantic 11 ± 5°C; mean ± SD) and variable areas inland (e.g. Novosibirsk in Siberia 1 ± 14°C). A premature *CMT2* stop codon located at 10 414 556 bp segregated in the RegMap collection with a minor allele frequency (MAF) of 0.05. This *CMT2$_{STOP}$* allele was in strong LD ($r^2 = 0.82$) with the leading SNP (Fig. 1B). The geographic distribution of the *CMT2$_{STOP}$* and *CMT2* wild-type (*CMT2$_{WT}$*) alleles in the RegMap collection shows that the *CMT2$_{WT}$* allele is almost exclusively found in accessions from areas with low temperature seasonality and that the *CMT2$_{STOP}$* allele exists across a broader range of climates. This result strongly suggest that this locus is involved in adaptation to temperature seasonality (Fig. 2A).

The *CMT2*/temperature seasonality association was also found in a separate dataset containing 665 re-sequenced accessions from the 1001-genomes project (http://1001genomes.org) [6,7,9-11] for which geographic origins were available (Supplementary Methods). In this more geographically diverse set (Fig. 2A), $CMT2_{STOP}$ was more common (MAF = 0.10). Two additional mutations were also identified on unique haplotypes ($r^2$ = 0.00) - one nonsense $CMT2_{STOP2}$ at 10 416 213 bp (MAF = 0.02) and a frameshift mutation at 10 414 640 bp (two accessions). Single locus association tests for $CMT2_{STOP2}$ and $CMT2_{STOP2}$ revealed similar genetic effects for both alleles (Fig. 1B) that were highly significant for the $CMT2_{STOP}$ allele (nominal $P = 1.1 \times 10^{-17}$) and non-significant for $CMT2_{STOP2}$ (nominal $P = 0.06$). The association test for $CMT2_{STOP2}$ was, however, underpowered due to the small sample-size ($n = 17$). Also, the existence of multiple, wide-spread mutations disrupting $CMT2$ suggest it as an evolutionary beneficial event.

$CMT2$ is a regulator of DNA methylation in *A. thaliana* [18]. Epigenetics, including chromatin and DNA methylation-based mechanisms, has been suggested as a potentially adaptive inheritance mechanism at the interface between genetic control and the environment [19]. DNA methylation defects leads to pleiotropic morphological changes in plants [20], including phenotypic plasticity [21]. Transcriptional reprogramming is also a central regulatory mechanism in temperature response [22]. The integrity of the RNA-directed DNA methylation (RdDM) pathway [23] is important for basal heat tolerance in *A. thaliana* [24]. *CMT2* and the *Arabidopsis* nucleosome remodeler *DDM1* is involved in an RdDM independent CHH methylation and plants with a mutated *CMT2* display a major loss of CHH methylation [18]. These results, together with the identification of three loss-of-function alleles across the natural *A. thaliana* accessions, suggest a role of an altered genome-wide methylation pattern, i.e. a *trans*- regulation of the epigenetic states of other obligatory epialleles [19], in adaptation to seasonal temperature variability.

We tested the epigenetic effect of $CMT2_{STOP}$ on genome-wide DNA methylation using 131 $CMT2_{WT}$ and 17 $CMT2_{STOP}$ accessions, for which MethylC-sequencing data was publicly available [11]. A methylome-wide association (MWA) analysis between $CMT2_{STOP}$ and the methylation-state at ~6 million single methylation polymorphisms (SMPs) identified ~3,000 methylome-wide significant associations (Supplementary Fig. 15). The multi-locus association patterns were visualized using "identity-by-methylation-state" (IBMS) matrices that quantify the total pairwise similarity in the methylation-patterns between accessions separately for all methylation types (CG, CHG and CHH). Fig. 3A shows the CHH-IBMS which illustrates that CHH-methylation is homogenous across the clear majority of these sites for the $CMT2_{WT}$ accessions. Interestingly, the methylation-pattern is more heterogenous among the $CMT2_{STOP}$ accessions, with a small overlap both among $CMT2_{STOP}$ accessions and between $CMT2_{STOP}$ and $CMT2_{WT}$ accessions, indicating a shared residual, non-$CMT2$ mediated CHH methylation. We confirmed that the differential methylation detected in the $CMT2_{STOP}$ and $CMT2_{WT}$ accessions is consistent with the effects of disrupting $CMT2$, by showing that the level of CHH-methylation across the MWA detected sites was significantly lower in four t-DNA insertion $cmt2$ knock-out [18] than in $CMT2_{WT}$ plants (Fig. 3B). No such difference was found for CHG and CG methylation sites (Fig. 3C,D).

A significant two-locus interaction between $CMT2_{STOP}$ and the leading SNP at the Beta-galactosidase 8 (*BGAL8*) locus was associated with temperature-seasonality (Supplementary Fig. 14A). When considered jointly in the RegMap collection [4], *BGAL8* and *CMT2* had significant marginal (nominal $P_{BGAL8} = 2.6 \times 10^{-26}$ and $P_{CMT2} = 4.8 \times 10^{-4}$) and interaction (nominal $P_{BGAL8 \times CMT2} = 1.0 \times 10^{-3}$) effects. In the 1001-genomes data [6,7,9-11], the effects were even more significant (nominal $P_{BGAL8} = 3.3 \times 10^{-29}$; $P_{CMT2} = 7.4 \times 10^{-4}$; $P_{BGAL8 \times CMT2} = 6.5 \times 10^{-5}$), most likely due to the larger number of minor-allele double homozygotes ($n_{RegMap} = 17$ v.s. $n_{1001-genomes} = 32$). The accessions carrying both alternative alleles are found in the regions with the most extreme seasonal variability in temperature (Supplementary Fig. 14B).

We identified several alleles associated with a broader range of climates across the native range of *A. thaliana*, suggesting that a genetically mediated plastic response might be of importance in climate adaptation. A mutated epigenetic mechanism involving *DDM1*-dependent, *CMT2* mediated CHH-methylation was strongly associated with adaptation to variability in seasonal temperatures, a finding suggesting that genetically determined epigenetic variability might contribute to a phenotypic plasticity of adaptive advantage in natural environments. In a parallel study, Dubin et al found that RdDM-based TE silencing is temperature sensitive and under control by *CMT2*, a finding that supports our finding of a likely involvement of *CMT2* variation in adaptation to temperature. The highly significant epistatic interaction between *CMT2* and *BGAL8* has no obvious molecular explanation. We interpret it as a possible reflection of the fact that fitness is determined jointly by multiple life-history traits that together contribute to survival in extreme environments. When multiple loci are beneficial in the same environment, this will then result in a statistical interaction in the association analysis. Traits such as seed dormancy, vegetative growth rate and flowering time follow latitudinal and longitudinal gradients, and joint evolution of these traits might therefore be adaptive [25]. Here, the *CMT2* differential methylation, together with the temperature-dependent expression of BGAL8 expression [26] and the general role of galactosidases in ripening and germination [27], might be a contributing factor to this complex pattern of multi-trait adaptation.

**Methods summary:**

We performed a genome-wide association (GWA) analysis for genetic variance heterogeneity between genotypes [12] for 13 climate variables across 214,553 SNPs in the RegMap *A. thaliana* collection [14]. A full linear mixed model utilizing a recently developed estimator for the genomic kinship [13] was implemented to correct for the population structure and control the rate of false-positives due to population stratification. Stringent quality control (QC) was used including a 5% minor allele frequency (MAF) cutoff in combination with genomic control [28] and model checking

[29]. The association between *CMT2* and temperature seasonality was replicated using the 665 accessions from the 1001-genomes project (http://1001genomes.org) [6,7,9-11] with available geographic origins. The methylome-wide association (MWA) analysis was conducted at 6,120,869 SMPs that were methylated in more than 5% of the 131 $CMT2_{WT}$ and 17 $CMT2_{STOP}$ accessions with MethylC-sequencing data [30] using the "qtscore" routine in the GenABEL package [31]. We screened for functional candidate mutations in high LD ($r^2 > 0.8$) with the leading SNP in the regions 100 kb up- and down-stream of the leading SNPs at the 16 significant loci using 728 available full-genome sequences from the 1001-genomes project [6-11]. Mutations were classified using the Ensembl Variant Effect Predictor [32], and functional effects of missense mutations were predicted using the PASE software [33].

**Full Methods** and any associated references are available in the online version of the paper.

**Acknowledgements:** Funded by a EURYI-award and SSF Future Research Leader Grant to Ö.C. We thank L. Andersson, L. Hennig, and. J. Lachowiec for helpful input. Also, providers of pre-publication sequence data within the 1001-genomes project are acknowledged for their efforts in creating this community resource, including Monsanto Company, the Weigel laboratory at the Max Planck Institute for Developmental Biology, the IGS of the Center for Biotechnology of the University of Bielefeld, the DOE Joint Genome Institute (JGI), the Joint BioEnergy Institute, the Nordborg laboratory of the Gregor Mendel Institute of Molecular Plant Biology, the Bergelson lab of the University of Chicago and the Ecker lab of the Salk Institute for Biological Studies, La Jolla, CA. Further, we also thank the contributors to the 19 genomes project, coordinated by the Mott group at the Wellcome Trust Sanger Institute, and the Zilberman at University of California at Berkely for sharing their earlier published data. Author contributions were: X.S. and Ö.C. conceived and designed the experiments, contributed to all analyses and wrote the paper. Ö.C. led and coordinated the study. X.S. developed the method for performing the genome-scan. S.F contributed to the analyses of the RNA-seq data. M.P. contributed to the replication association analysis. Z.S. contributed to the functional analyses of genetic polymorphisms. S.F., M.P. and Z.S. commented on the manuscript.

*Supporting Online Material:*

Materials and Methods

Supplementary Text

Supplementary Figures 1 - 15

Supplementary Tables 1 - 2

References 34-39

**Figures:**

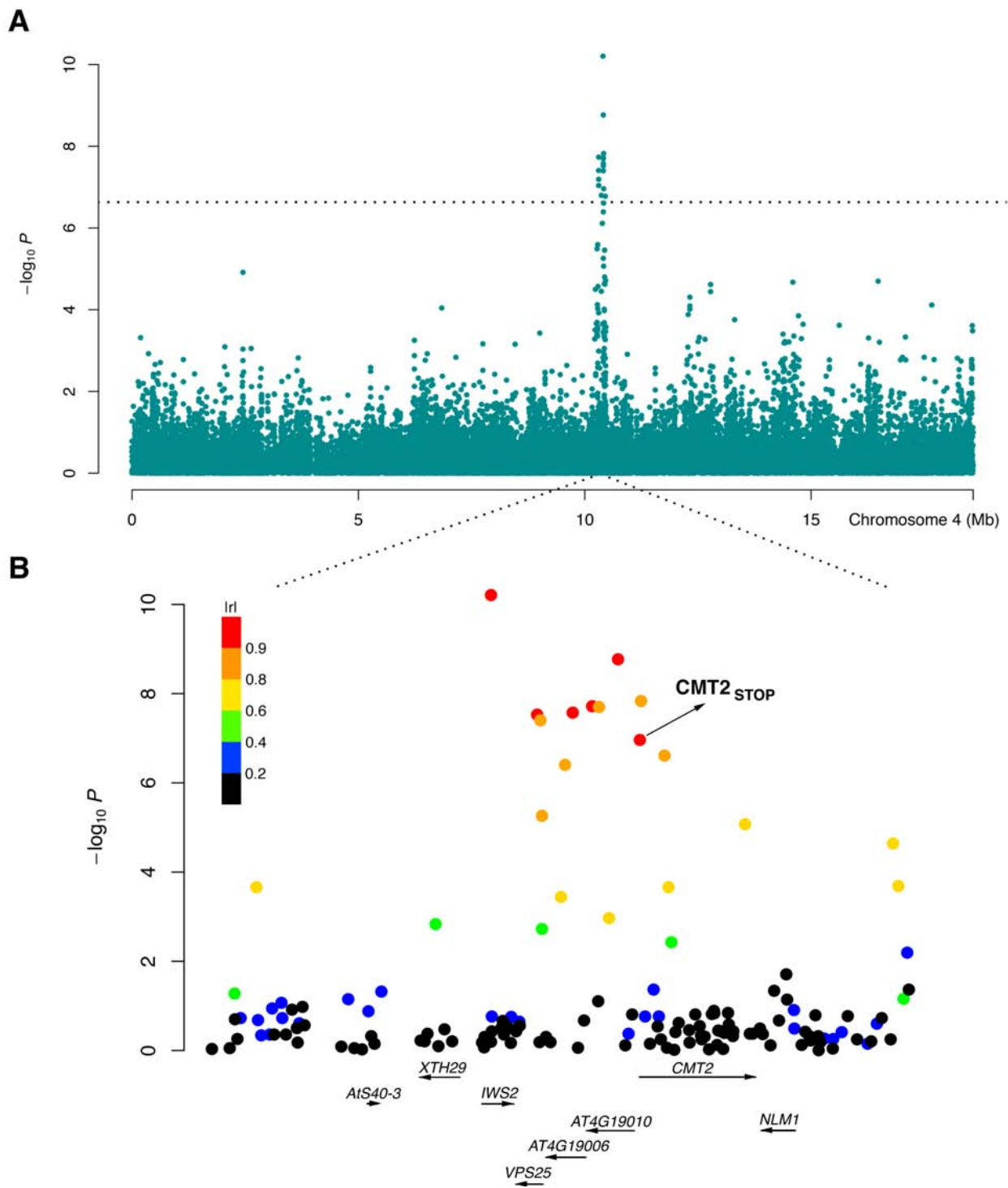

**Figure 1.** An LD block associated with temperature seasonality contains *CMT2*. A genome-wide significant variance-heterogeneity association signal was identified for temperature seasonality in the RegMap collection of natural *Arabidopsis thaliana* accessions [4]. The peak on chromosome 4 around 10 Mb **(A)** mapped to a haplotype block **(B)** containing a nonsense mutation (CMT2$_{STOP}$) early in the first exon of the Chromomethylase 2 (*CMT2*) gene. Color coding based on |r| (the

absolute value of the correlation coefficient) as a measure of LD between each SNP in the region and the leading SNP in the association analysis.

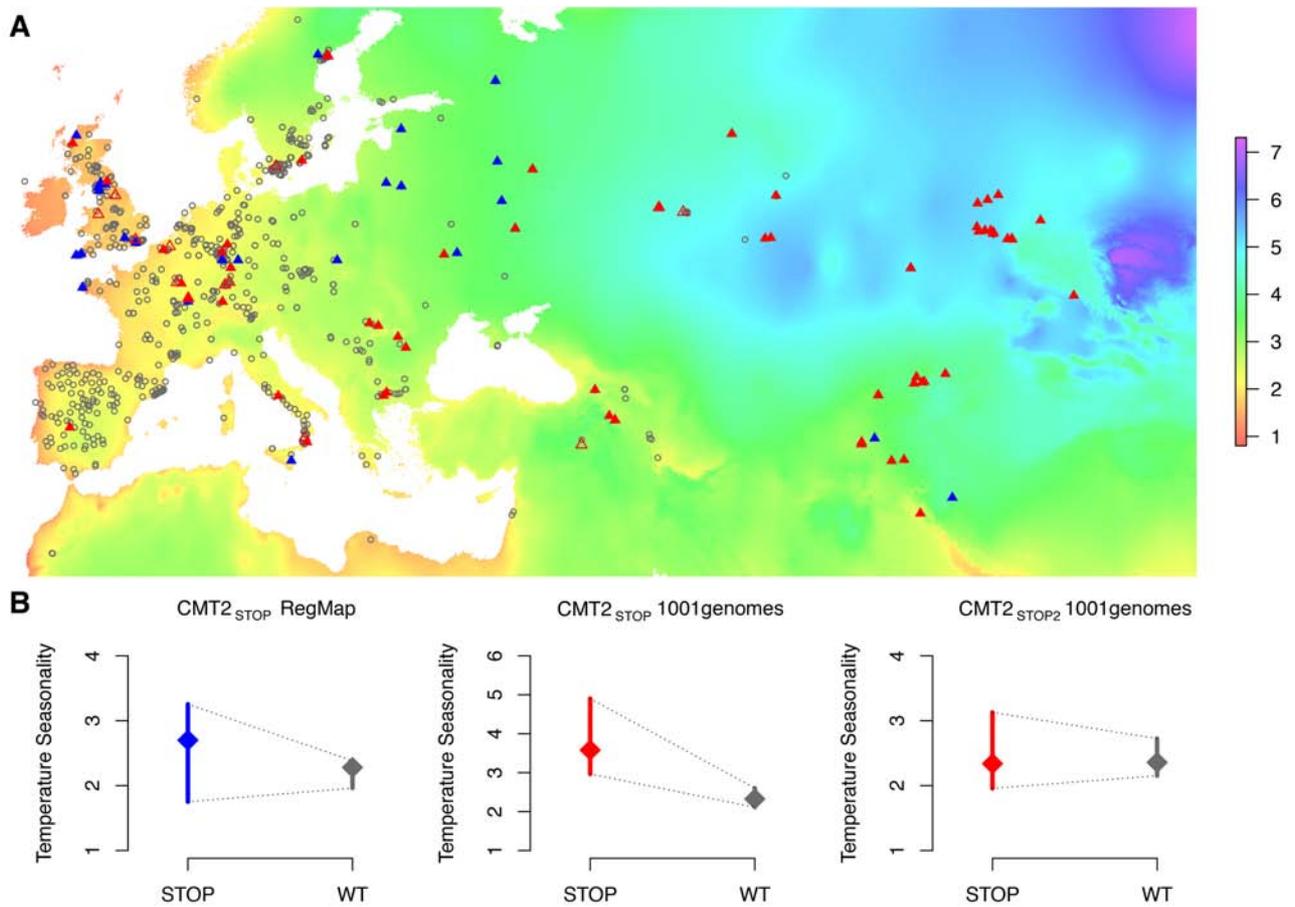

**Figure 2.** Geographic distribution of, and heterogenous variance for, three *CMT2* alleles in two collections of *A. thaliana* accessions. The geographic distributions **(A)** of the wild-type (CMT2$_{WT}$; gray circles) and two nonsense alleles (CMT2$_{STOP}$/CMT2$_{STOP2}$; filled/open triangles) in the *CMT2* gene that illustrates a clustering of CMT2$_{WT}$ alleles in less variable regions and a greater dispersion of the nonsense alleles across different climates both in the RegMap [4] (blue) and the 1001-genomes [6](red) *A. thaliana* collections. The resulting variance-heterogeneity in temperature seasonality between genotypes is highly significant, as illustrated by the quantile plots in **(B)** where the median is indicated by a diamond and a bar representing the 25% to 75% quantile range. The color scale indicate the level of temperature seasonality across the map. The colorkey in **(A)** represent the temperature seasonality values, given as the standard-deviation in % of the mean temperature (K).

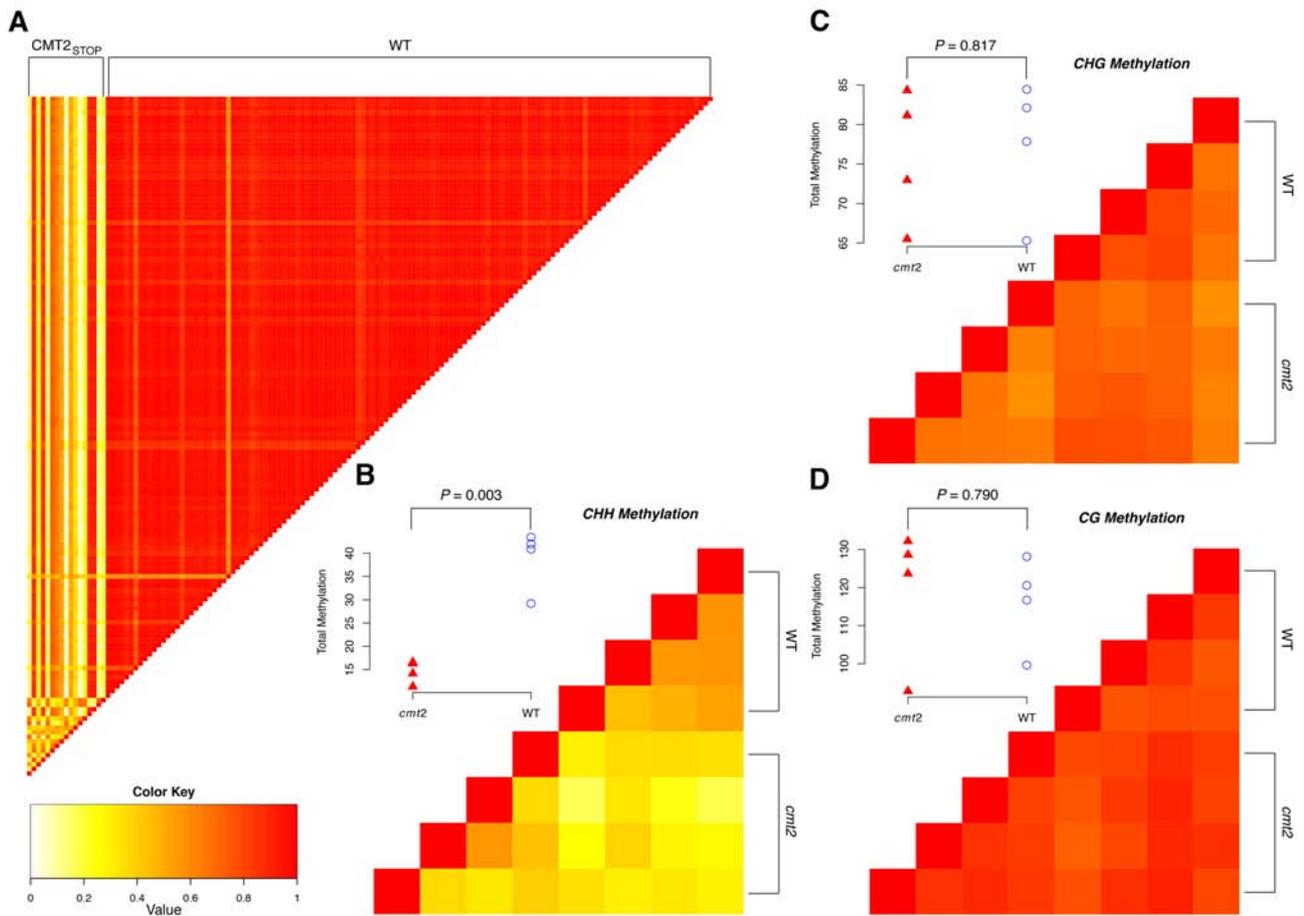

**Figure 3.** Genome-wide methylation patterns for different *CMT2* genotypes. The genome-wide CHH-methylation pattern is similar across the 131 *A. thaliana* CMT2$_{WT}$ **(A)**, but not so among the 17 accessions carrying the loss-of-function (CMT2$_{STOP}$) allele or between the CMT2$_{WT}$ and CMT2$_{STOP}$ accessions. The most divergently CHH methylated sites between natural CMT2$_{WT}$ and STOP$_{CMT2}$ accessions are also differentially methylated between CMT2$_{WT}$ and *cmt2* plants, illustrated here by the degree of methylation sharing at these sites amongst four CMT2$_{WT}$ and four *cmt2* mutants (*cmt2-3*, *cmt2-4*, *cmt2-5*, *cmt2-6*) [18] **(B)**. No such differential methylation was found in neither for CHG **(C)** nor CG **(D)** sites. The color key represent identity-by-methylation-state (IBMS) values in (**A**), and correlation coefficients of methylation scores in (**B**, **C**, **D**).

**Table:**

Table 1. Loci with genome-significant, non-additive effects on climate adaptation and a functional analysis of nearby genes ($r^2 > 0.8$) containing missense or nonsense mutations.

| Trait | Leading SNP | | | | Selected Candidate Genes | | | Mutant Analysis[1] | |
|---|---|---|---|---|---|---|---|---|---|
| | Chrom | Pos (bp) | MAF[2] | P-value[3] | Effect[4] | Gene name | Locus | # Mut[5] | PASE | MSA |
| Temperature seasonality | | | | | | | | | | |
| | 2[a] | 12 169 701 | 0.08 | 2.0E-08 | 0.53±0.08 | BGAL8 | AT2G28470 | 4 | 0.36[6] | 0.72[6] |
| | 4 | 10 406 018 | 0.05 | 6.1E-11 | 0.51±0.06 | IWS2 | AT4G19000 | 1 | 0.21 | 0.35 |
| | | | | | | CMT2 | AT4G19020 | 7 | STOP | STOP |
| Maximum temperature in the warmest month | | | | | | | | | | |
| | 1 | 6 936 457 | 0.05 | 1.9E-07 | 0.34±0.07 | | AT1G19990 | 1 | 0.64 | 0.2 |
| Minimum temperature in the coldest month | | | | | | | | | | |
| | 2[b] | 18 620 697 | 0.08 | 3.5E-08 | 0.33±0.05 | | | | | |
| | 2[c] | 19 397 389 | 0.05 | 5.0E-08 | 0.38±0.06 | | | | | |
| | 5 | 14 067 526 | 0.07 | 4.2E-08 | 0.33±0.05 | | AT5G35930 | 1 | 0.30 | 0.05 |
| | 5[d] | 18 397 418 | 0.11 | 1.2E-07 | 0.28±0.05 | | | | | |
| Number of consecutive cold days | | | | | | | | | | |
| | 2[b] | 18 620 697 | 0.08 | 1.7E-07 | 0.33±0.05 | | | | | |
| | 2[c] | 19 397 389 | 0.05 | 7.2E-08 | 0.39±0.06 | | | | | |
| | 5 | 7 492 277 | 0.08 | 4.3E-09 | 0.38±0.06 | | AT5G22560 | 4 | 0.63 | 0.11 |
| | 5[d] | 18 397 418 | 0.11 | 1.9E-07 | 0.29±0.05 | | | | | |
| Day length in spring | | | | | | | | | | |
| | 2[a] | 12 169 701 | 0.08 | 2.0E-07 | 0.46±0.08 | BGAL8 | AT2G28470 | 4 | 0.36 | 0.72 |
| | 3 | 12 642 006 | 0.07 | 9.4E-08 | 0.29±0.05 | | | | | |
| | 4[e] | 14 788 320 | 0.08 | 2.2E-08 | 0.39±0.06 | VEL1 | AT4G30200 | 2 | 0.26 | 0.06 |
| Relative humidity in spring | | | | | | | | | | |
| | 3 | 1 816 353 | 0.07 | 1.2E-08 | 0.39±0.06 | | | | | |
| | 4[e] | 14 834 441 | 0.06 | 6.4E-08 | 0.49±0.08 | VEL1 | AT4G30200 | 2 | 0.26 | 0.06 |
| | | | | | 0.43±0.07 | XTH19 | AT4G30290 | 1 | 0.14 | 0.49 |
| | 5 | 8 380 640 | 0.07 | 6.3E-08 | | | | | | |
| Length of the growing season | | | | | | | | | | |
| | 3 | 576 148 | 0.08 | 1.4E-07 | 0.27±0.04 | | | | | |
| Number of consecutive frost-free days | | | | | | | | | | |
| | 1 | 953 031 | 0.24 | 8.0E-08 | 0.25±0.04 | SOM | AT1G03790 | 3 | 0.25 | 0.55 |
| | 1 | 6 463 065 | 0.08 | 1.9E-07 | 0.33±0.06 | | | | | |
| | 2 | 9 904 076 | 0.22 | 2.2E-08 | 0.22±0.04 | | | | | |

[a,b,c,d,e]Loci affecting affect multiple traits; [1]The predicted functional effect score for the strongest mis-sense mutation in the gene based on amino-acid physiochemical properties (PASE) and evolutionary conservation (MSA) [33]; [2]Locus contains two mis-sense mutations with equally strong predicted effects; [2]MAF: Minor Allele Frequency; [3]P-value: significance after genomic-control from a linear regression analysis of squared z-scores accounting for population stratification. [4]Effect: Standardized genetic effect on adaptability (Chi-square distributed) ± standard error (unit: phenotypic standard deviations); [5]#Mut: number mis- and non-sense mutations in the gene in the

1001-genomes dataset [6]. [6]Locus contains two missense mutations with equally strong predicted effects

# Supplementary Information for

# Natural *CMT2* variation is associated with genome-wide methylation changes and temperature adaption

Xia Shen, Simon Forsberg, Mats Pettersson, Zheya Sheng and Örjan Carlborg*

*To whom correspondence should be addressed. E-mail: orjan.carlborg@slu.se

**This PDF file includes:**

Materials and Methods
Supplementary Text
Supplementary Fig. 1-15
Supplementary Table 1-2
References 34-39

# Materials and Methods

## Climate data and genotyped *Arabidopsis thaliana* accessions

The climate phenotypes and *A. thaliana* genotype data that we re-analyzed were obtained from [4]. 13 climate variables and genotypes of 214,553 single nucleotide polymorphisms (SNPs) for 948 accessions were available at: http://bergelson.uchicago.edu/regmap-data/climate-genome-scan. The climate variables used in the analyses were: aridity, number of consecutive cold days (below 4 degrees Celsius), number of consecutive frost-free days, daylength in the spring, growing season length, maximum temperature in the warmest month, minimum temperature in the coldest month, temperature seasonality, photosynthetically active radiation, precipitation in the wettest month, precipitation in the driest month, precipitation seasonlity, and relative humidity in the spring. No squared pairwise Pearson's correlation coefficients between the phenotypes were greater than 0.8 (Fig. S7 of [4]).

## Genome-wide association analysis to identify adaptability loci in the RegMap collection

GWAS datasets based on natural collections of *A. thaliana* accessions, such as the RegMap collection, are often genetically stratified. This is primarily due to the close relationships between accessions sampled at nearby locations. Furthermore, as the climate measurements used as phenotypes for the accessions are values representative for the sampling locations of the individual accessions, these measurements will be confounded with the general genetic relationship [12]. Unless properly controlled for, this confounding might lead to excessive false-positive signals in the association analysis; this as the differences in allele-frequencies between loci in locations that differ in climate, and at the same time are geographically distant, will create an association between the genotype and the trait. However, this association could also be due to other forces than selection. In traditional GWAS analyses, mixed-model based approaches are commonly used to control for population-stratification. The downside of this approach is that it, in practice, will remove many true genetic signals coming from local adaptation due to the inherent confounding between local genotype and adaptive phenotype. Instead, the primary signals from such analyses will be due to effects of alleles that exist in, and have similar effects across, the entire studied population. Although studies into the contributions of genetic variance-heterogeneity to the phenotypic variability in complex traits is generally considered to be a novel and useful approach with great potential [34], some concerns have been raised regarding the potential statistical pitfalls of large-scale applications of such analyses [29]. Here, we have developed, and used, a new linear mixed models based approach that addresses these initial concerns and shown that it is possible to infer statistically robust results of genetically regulated phenotypic variability in GWA data from natural populations.

## Statistical modeling in genome-wide scans for adaptability

The climate data at the geographical origins of the *A. thaliana* accessions were treated as phenotypic responses. Each climate phenotype vector **y** for all the accessions were normalized via an inverse-Gaussian transformation. The squared normalized measurement $z_i = y_i^2$ of accession $i$ is modeled by the following linear mixed model to test for an association with climate adaptability (i.e. a greater plasticity to the range of the environmental condition):

$$z_i = \mu + \beta x_i + g_i + e_i$$

where $\mu$ is an intercept, $x_i$ the SNP genotype for accession $i$, $\beta$ the genetic SNP effect, $\mathbf{g} \sim MVN(\mathbf{0}, \mathbf{G}^* \sigma_g^2)$ the polygenic effects and $\mathbf{e} \sim MVN(\mathbf{0}, \mathbf{I}\sigma_e^2)$ the residuals. $x_i$ is coded 0 and 2 for the two homozygotes (inbred lines). The genomic kinship matrix $\mathbf{G}^*$ is constructed via the whole-genome generalized ridge regression method HEM (heteroscedastic effects model) [14] as $\mathbf{G}^* = \mathbf{ZWZ}'$, where **Z** is a number of individuals by number of SNPs matrix of genotypes standardized by the allele frequencies. **W** is a diagonal matrix with element $w_{jj} = \hat{b}_j/(1 - h_{jj})$ for the $j$-th SNP, where $\hat{b}_j$ is the SNP-BLUP (SNP-Best Linear Unbiased Prediction) effect estimate for the $j$-th SNP from a whole-genome ridge regression, and $h_{jj}$ is the hat-value for the $j$-th SNP. Quantities in **W** can be directly calculated using the **bigRR** package [14]: http://cran.r-project.org/web/packages/bigRR/index.html. An example R source code for performing the analysis is provided below.

The advantage of using the HEM genomic kinship matrix $\mathbf{G}^*$, rather than an ordinary genomic kinship matrix $\mathbf{G} = \mathbf{ZZ}'$, is that HEM is a significant improvement of the ridge regression based (SNP-BLUP) in terms of the estimation of genetic effects [14]. Due to this, the $\mathbf{G}^*$ accurately represents the relatedness between accessions and also accounts for the genetic effects of the SNPs on the phenotype.

## Example R source code for calculating HEM genomic kinship matrix

Here, we use the example data in the **bigRR** package: http://cran.r-project.org/web/packages/bigRR/ to illustrate how an ordinary identity-by-state (IBS) kinship matrix can be update to a HEM genomic kinship matrix. The full theoretical details on this procedure are provided in [11].

```r
# load the bigRR package
require(bigRR)

# load the example data
data(Arabidopsis)
X <- matrix(1, length(y), 1)
Z <- scale(Z)

# fitting SNP-BLUP, i.e. a ridge regression on all the markers across the genome
SNP.BLUP <- bigRR(y = y, X = X, Z = Z, family = binomial(link = 'logit'))

# calculate HEM (heteroscedastic effects model) genomic kinship matrix
w <- as.numeric(SNP.BLUP$u^2/(1 - SNP.BLUP$leverage))
wZt <- sqrt(w)*t(Z)
G <- crossprod(wZt)

# visualization HEM genomic kinship matrix G
image(G)
```

## Phenotype preparation in R

We present our original routine for phenotyping temperature seasonality in Euro-Asia as an example. The data downloaded from http://www.worldclim.org/ were processed using the following code to obtain an object readable by the **raster** package: http://cran.r-project.org/web/packages/raster/.

```r
# load the original data files
bil_files <- grep(".bil", dir("tmean_30s_bil Folder/"), value = T)
bil_file_order <- as.numeric(sub(pattern = ".+_([0-9]+).bil",
                  x = bil_files, replacement = "\\1"))
bil_files <- bil_files[order(bil_file_order)]

# create rasters
WorldClim_stack <- stack()
```

```
for (bil_file in bil_files){
r <- raster(paste("tmean_30s_bil Folder/",bil_file, sep = ""))
WorldClim_stack <- addLayer(WorldClim_stack, r)
}

# temperature seasonality calculation
r_mean <- calc(WorldClim_stack, mean)
save(r_mean, file = "WorldClim_mean.Rdata")
writeRaster(r_mean, file = "WorldClim_mean.raster")
r_sd <- calc(WorldClim_stack, sd)
save(r_sd, file = "WorldClim_sd.Rdata")
writeRaster(r_sd, filename = "WorldClim_sd.raster")
r_mean_corr <- r_mean/10 + 273.15
save(r_mean_corr, file = "WorldClim_mean_corr.Rdata")
r_coeff_var <- 100*(r_sd/10)/r_mean_corr

# output to a raster object
writeRaster(r_coeff_var, file = "WorldClim_coeff_var.raster")

# phenotyping at given coordinates
# (LONGITUDE and LATITUDE already loaded)
require(raster)
world_temp_seas <- raster('WorldClim_coeff_var')
temp_seas <- raster::extract(world_bio5, cbind(LONGITUDE, LATITUDE))
```

### Testing and quality control for association with climate adaptability

The test statistic for the SNP effect $\beta$ is constructed as the score statistic by [35]:

$$T^2 = \frac{(\tilde{\mathbf{x}}'\mathbf{G}^{*-1}\tilde{\mathbf{z}})^2}{\tilde{\mathbf{x}}'\mathbf{G}^{*-1}\tilde{\mathbf{x}}}$$

implemented in the **GenABEL** package [31], where $\tilde{\mathbf{x}} = \mathbf{x} - E[\mathbf{x}]$ and $\tilde{\mathbf{z}} = \mathbf{z} - E[\mathbf{z}]$. The $T^2$ statistic has an asymptotic $\chi^2$ distribution with 1 degree of freedom. This was then used to perform genomic control (GC) [31] of the genome-wide association results under the null hypothesis that no SNP has an effect on the climate phenotype. SNPs with minor allele frequency (MAF) less than 0.05 were excluded from the analysis. A 5% Bonferroni-corrected significance threshold was applied. As suggested by [29], the significant SNPs were also analyzed using a Gamma generalized linear model to exclude positive findings that might be due to low allele frequencies of the high-variance SNP.

## Functional analysis of polymorphisms in loci with significant genome-wide associations to climate

All the loci that showed genome-wide significance in the association study was further characterized using the genome sequences of 728 accessions sequenced as part of the 1001-genomes project (http://1001genomes.org). Mutations within a ±100Kb interval of each leading SNP, and that are in high linkage disequilibrium (LD) with the leading SNP ($r^2 > 0.8$), were reported (Supplementary Table 1). The consequences of the identified polymorphisms were predicted using the Ensembl variant effect predictor [32] and their putative effects on the resulting protein estimated using the PASE (Prediction of Amino acid Substitution Effects) tool [33].

## Methylome-wide association analysis and validation of $CMT2_{STOP}$ genotypes

A methylome-wide association (MWA) analysis was conducted to the $CMT2_{STOP}$ genotypes at 43,182,344 scored single methylation polymorphisms (SMPs) across the genome. 131 $CMT2_{WT}$ and 17 $CMT2_{STOP}$ accessions, for which MethylC-sequencing data was publicly available at http://www.ncbi.nlm.nih.gov/geo/query/acc.cgi?acc=GSE43857 [11], were included in the analysis. Sites that were methylated at a frequency less than 0.05 among the accessions were removed from the analysis, resulting in 6,120,869 methylation sites to be tested across the genome. In this set, we tested for an association between the $CMT2_{STOP}$ genotype and the methylation state at each of the 6,120,869 SMPs using the "qtscore" routine in the **GenABEL** package [31].

In total, 3,096 methylome-wide SMPs were significant at a Bonferroni-corrected significance threshold for 6,120,869 tests. Divided according to the type of methylated sites, they corresponded to 879 CHH, 1162 CG and or 731 CHG SMPs. To visualize the pairwise similarity between accessions at these sites, we computed an identity-by-methylation-state (IBMS) matrix using all the significant CHH sites (Fig. 3A). To validate that the high degree of shared methylated sites was a useful predictor for the $CMT2_{STOP}$ genotype, we downloaded data from an independent experiment [18] that contained methylome data on four $CMT2$ knockouts and four WT samples (GSM1083504, GSM1083505, GSM1083506,

GSM1014134, GSM1014135, GSM1093622 and GSM1093629 from http://www.ncbi.nlm.nih.gov/geo/query/acc.cgi?acc=GSE41302GSM1014115). In total, the data from [18] contained scores for 718/213/262 of the differentially methylated CHH/CHG/CG sites. The level of CHH/CHG/CG methylation was scored in each of the eight samples as the sum of the methylation levels across all these CHH/CHG/CG sites. The respective methylation-levels for all samples are provided in Figure 3B/C/D and a t-test shows that the methylation-level was significantly different between lines with dysfunctional and WT *CMT2* for CHH sites, but not for CHG or CG sites. Together these results clearly shows that the *CMT2$_{STOP}$* accessions carry a mutated *CMT2* allele.

Our IBMS results (Fig. 3A) indicated that four of the 17 *CMT2$_{STOP}$* accessions (En-D, Fi-0, Stw-0 & Vind-1) displayed a CHH-methylation pattern across the differentially methylated sites that was closer to the *CMT2$_{WT}$* phenotype. Interestingly, an evaluation of the *CMT2* mRNA abundance in these accessions using data from [11] showed that these lines also had higher transcript levels than other *CMT2$_{STOP}$* accessions. Although *CMT2$_{STOP}$* is a strong candidate mutation to cause the mutant *CMT2* methylation phenotype, it is not possible to rule out that it is only in strong LD with an alternative causative variant or that other mechanisms are involved that causes the obligatory epialleles across the genome to be reverted by other compensatory mechanisms. Regardless, the results clearly show that the differential CHH-methylation phenotype is caused by a loss-of-function *CMT2* allele.

## Other Results & Further Discussions

### *CMT2* is a potential target for the nonsense-mediated RNA decay (NMD) pathway

To further explore the potential mechanism underlying the observed effect of *CMT2*, as well as the heterogeneity within the group of mutant accessions, we also studied the level of mRNA in the plants. The motivation for this evaluation was that the *CMT2$_{STOP}$* allele will produce an mRNA with a premature translation termination codon, which makes it a likely target for the nonsense-mediated mRNA decay (NMD) pathway. Thus, the expectation is that accessions carrying this allele would have lower transcript abundance than the wild-type. For this study, we used data from two studies that contained data both on the genotype for *CMT2* as well as RNA-seq data for the same lines. First, we analyzed the 19 genomes project data [36] that contained full genome sequences and transcriptomes for 19 *A. thaliana* accessions, 2 of which (Ct-1 and Kn-0) are part of the RegMap panel and carries the *CMT2$_{STOP}$* mutation according to their 250k SNP-chip genotypes [15]. Utilizing data from both the biological replicates of seedling mRNA, the difference in mRNA abundance was highly significant between mutant and wild type accessions ($P = 6.9 \times 10^{-5}$), with a higher expression in the wild-type. A similar analysis was done using a larger data set from [11], with *CMT2* genotypes obtained from SNPs called from whole-genome re-sequencing data and RNA-seq data was obtained from analyses of leaf tissue in 14 *CMT2$_{STOP}$* and 92 *CMT2$_{WT}$* accessions. Here, the average mRNA abundance was higher for the wild-type accessions, but the difference was not significant in the complete dataset ($P = 0.14$). However, the mRNA levels were significantly higher among the four mutant accessions that displayed a methylation pattern resembling that of the wild-type in the analysis above (t-test; $P = 0.01$) and when those lines were removed from the comparison, the levels of mRNA was significantly higher in the wild-type accessions than in the ten remaining mutants (t-test; $P = 0.02$). These results indicate that *CMT2* mRNA levels are influenced by the genotype and that it is connected to the methylation state in the plant, but provide no conclusive evidence on the functional connection between the two.

## *VEL1* and adaptation to day length

A thaliana is a facultative photoperiodic flowering plant and hence non-inductive photoperiods will delay, but not abolish, flowering. The genetic control of this phenotypic plasticity is thus an adaptive trait. A significant association was detected near two genes, *VEL1* and *XTH19*, containing two and one non-synonymous amino acid substitutions, respectively (Table 1). The major allele was dominant in short-day regions, whereas the alternative allele was more plastic in relation to day-length. *XTH19* has been implied as a regulator of shade avoidance [37], but information about its potential involvement in regulation of photoperiodic length is lacking. *VEL1*, regulates the epigenetic silencing of genes in the *FLC*-pathway in response to vernalization [38] and photoperiod length [39] resulting in an acceleration of flowering under non-inductive photoperiods. A feasible explanation for the existence of an adaptability *VEL1*-allele could thus be that accelerated flowering is beneficial under short-day conditions, but that also lack of accelerated flowering is allowed. In long-daytime areas, however, accelerated flowering might be detrimental as day-length follows a latitudinal cline, where early flowering might be detrimental in northern areas where accelerated flowering when the day-length is short could lead to excessive exposure to cold temperatures in the early spring and hence a lower fitness.

### a: Phenotypic and *p*-value distributions.

Top-left: phenotypic distribution; Top-right: $-\log_{10}p$-values after genomic control (GC) against minor allele frequencies (MAF); Bottom panels: Quantile-quantile plots of *p*-values and $-\log_{10}p$-values before (blue) and after (green) GC.

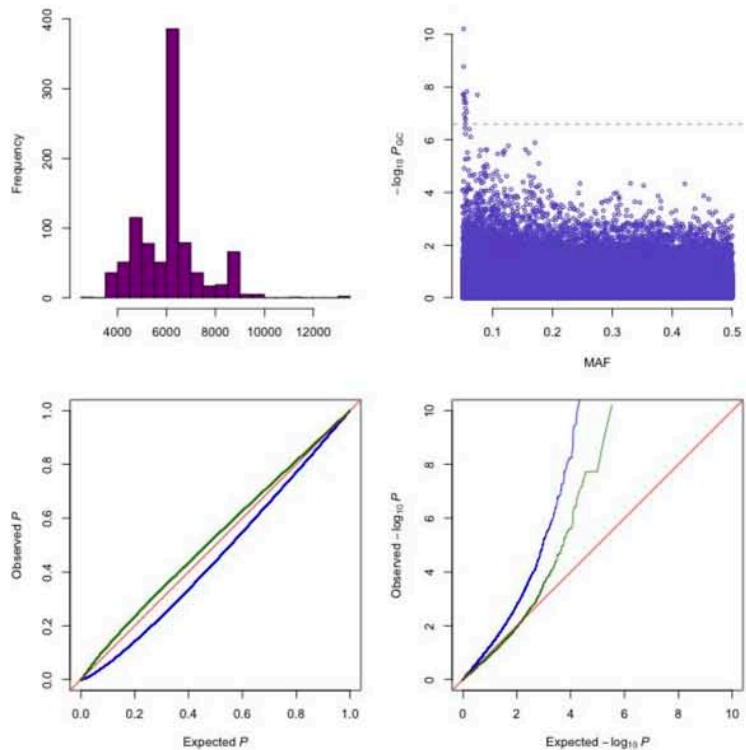

### b: Genome-wide association mapping for climate adaptability.

The plotted $-\log_{10}p$-values are genomic controlled. Markers with minor allele frequencies less than 5% are removed. Chromosomes are distinguished by colors. The Bonferroni-corrected significance threshold is marked by the horizontal line.

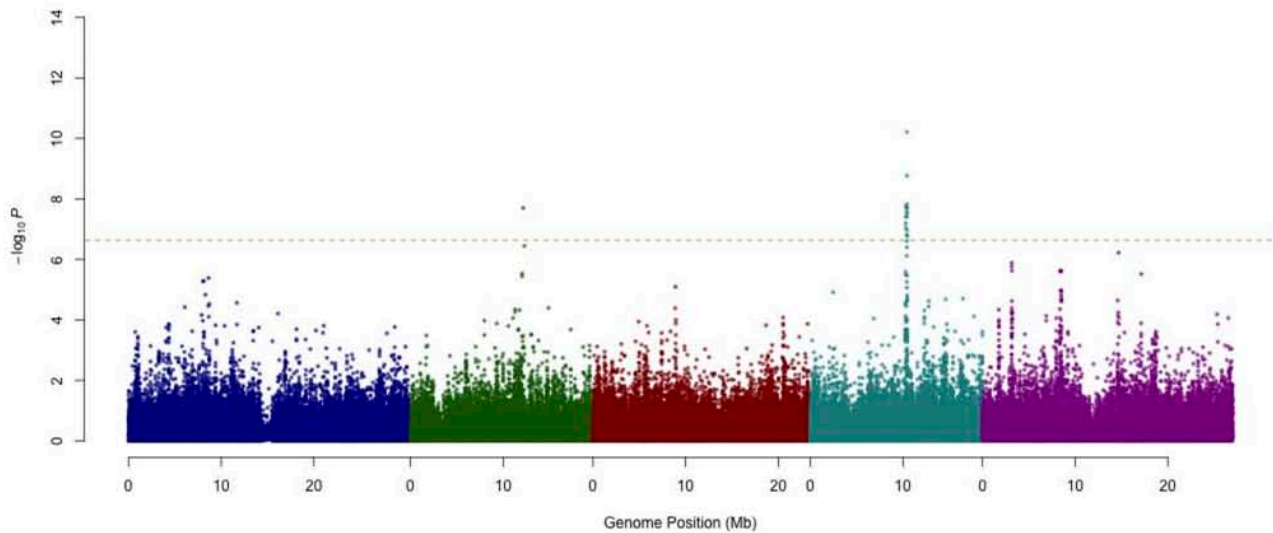

**Supplementary Figure 1** - Summary of results for temperature seasonality.

### a: Phenotypic and *p*-value distributions.

Top-left: phenotypic distribution; Top-right: $-\log_{10}p$-values after genomic control (GC) against minor allele frequencies (MAF); Bottom panels: Quantile-quantile plots of *p*-values and $-\log_{10}p$-values before (blue) and after (green) GC.

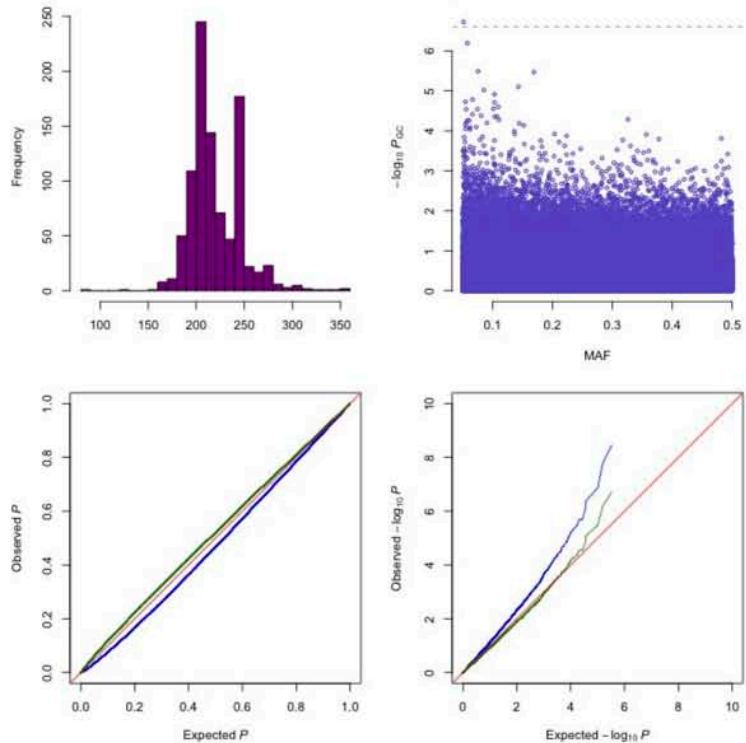

### b: Genome-wide association mapping for climate adaptability.

The plotted $-\log_{10}p$-values are genomic controlled. Markers with minor allele frequencies less than 5% are removed. Chromosomes are distinguished by colors. The Bonferroni-corrected significance threshold is marked by the horizontal line.

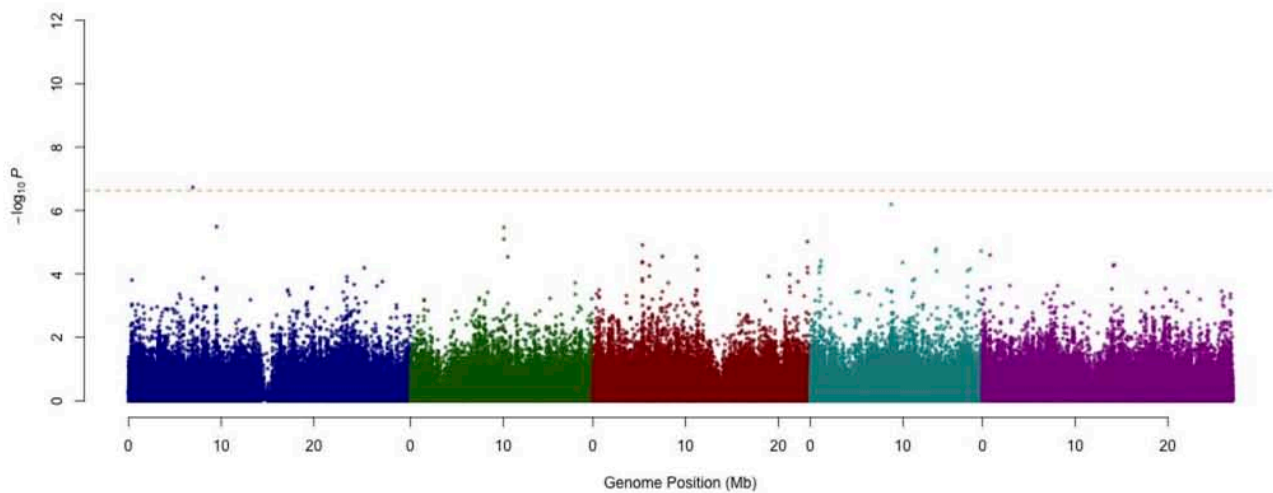

**Supplementary Figure 2** - Summary of results for maximum temperature in the warmest month.

### a: Phenotypic and *p*-value distributions.

Top-left: phenotypic distribution; Top-right: $-\log_{10}p$-values after genomic control (GC) against minor allele frequencies (MAF); Bottom panels: Quantile-quantile plots of *p*-values and $-\log_{10}p$-values before (blue) and after (green) GC.

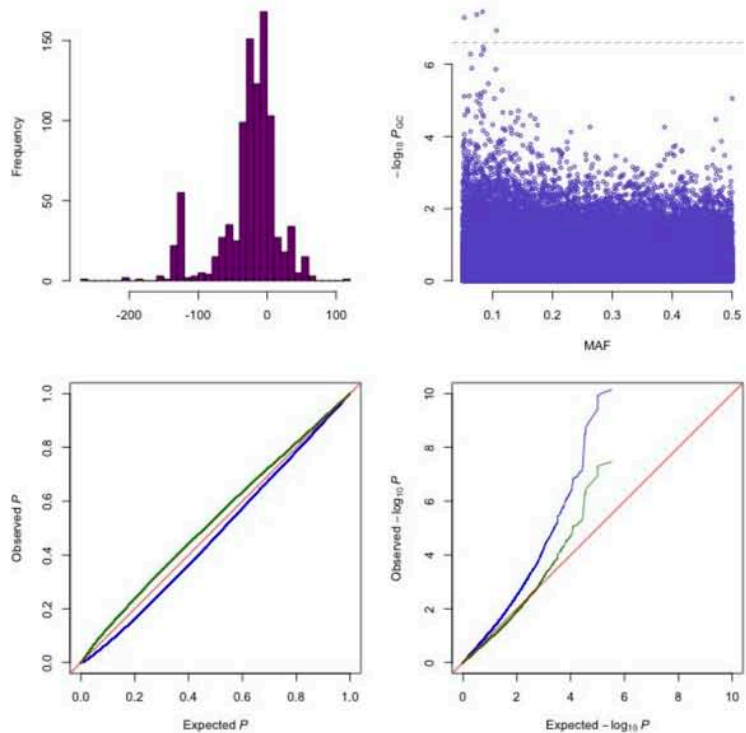

### b: Genome-wide association mapping for climate adaptability.

The plotted $-\log_{10}p$-values are genomic controlled. Markers with minor allele frequencies less than 5% are removed. Chromosomes are distinguished by colors. The Bonferroni-corrected significance threshold is marked by the horizontal line.

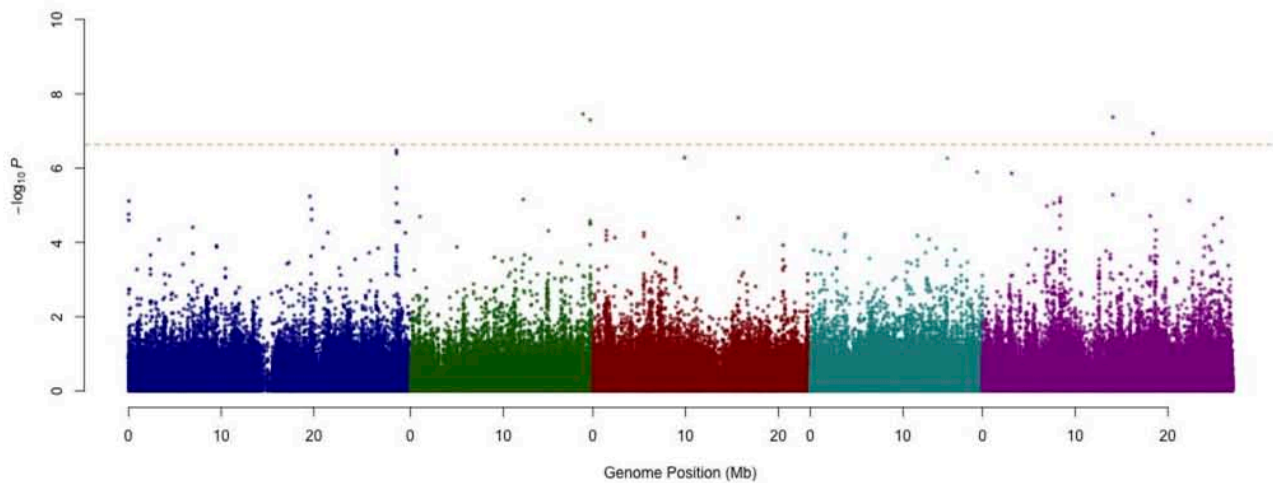

**Supplementary Figure 3** - Summary of results for minimum temperature in the coldest month.

### a: Phenotypic and *p*-value distributions.

Top-left: phenotypic distribution; Top-right: $-\log_{10}p$-values after genomic control (GC) against minor allele frequencies (MAF); Bottom panels: Quantile-quantile plots of *p*-values and $-\log_{10}p$-values before (blue) and after (green) GC.

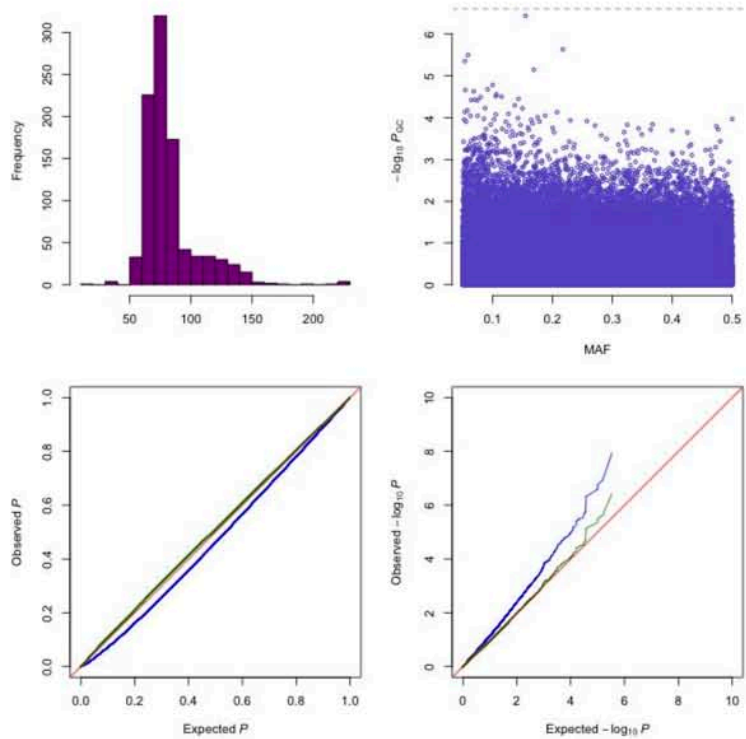

### b: Genome-wide association mapping for climate adaptability.

The plotted $-\log_{10}p$-values are genomic controlled. Markers with minor allele frequencies less than 5% are removed. Chromosomes are distinguished by colors. The Bonferroni-corrected significance threshold is marked by the horizontal line.

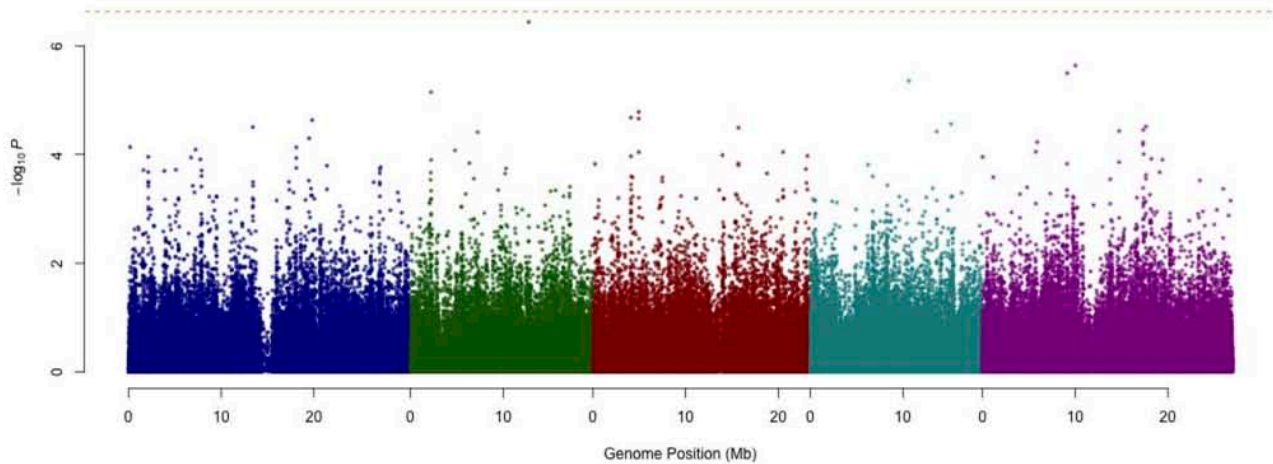

**Supplementary Figure 4** - Summary of results for precipitation in the wettest month.

### a: Phenotypic and *p*-value distributions.

Top-left: phenotypic distribution; Top-right: -$\log_{10}p$-values after genomic control (GC) against minor allele frequencies (MAF); Bottom panels: Quantile-quantile plots of *p*-values and -$\log_{10}p$-values before (blue) and after (green) GC.

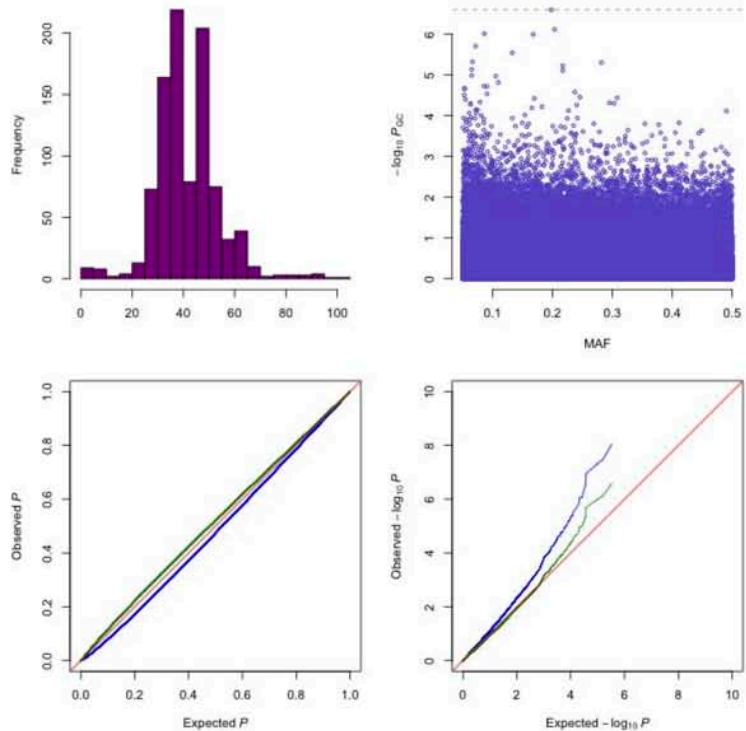

### b: Genome-wide association mapping for climate adaptability.

The plotted -$\log_{10}p$-values are genomic controlled. Markers with minor allele frequencies less than 5% are removed. Chromosomes are distinguished by colors. The Bonferroni-corrected significance threshold is marked by the horizontal line.

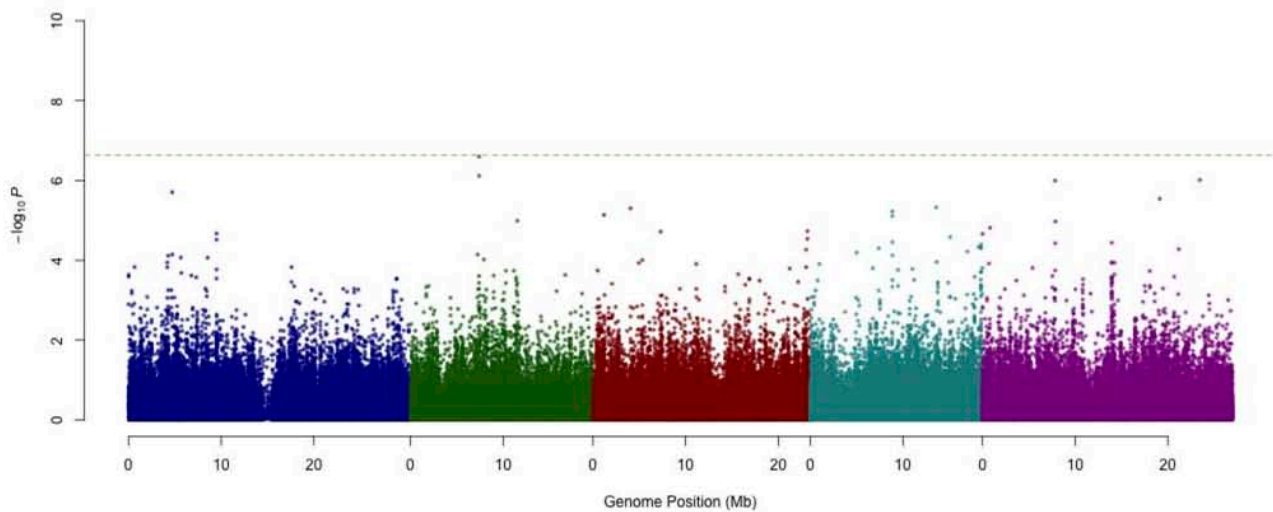

**Supplementary Figure 5** - Summary of results for precipitation in the driest month.

### a: Phenotypic and *p*-value distributions.

Top-left: phenotypic distribution; Top-right: $-\log_{10}p$-values after genomic control (GC) against minor allele frequencies (MAF); Bottom panels: Quantile-quantile plots of *p*-values and $-\log_{10}p$-values before (blue) and after (green) GC.

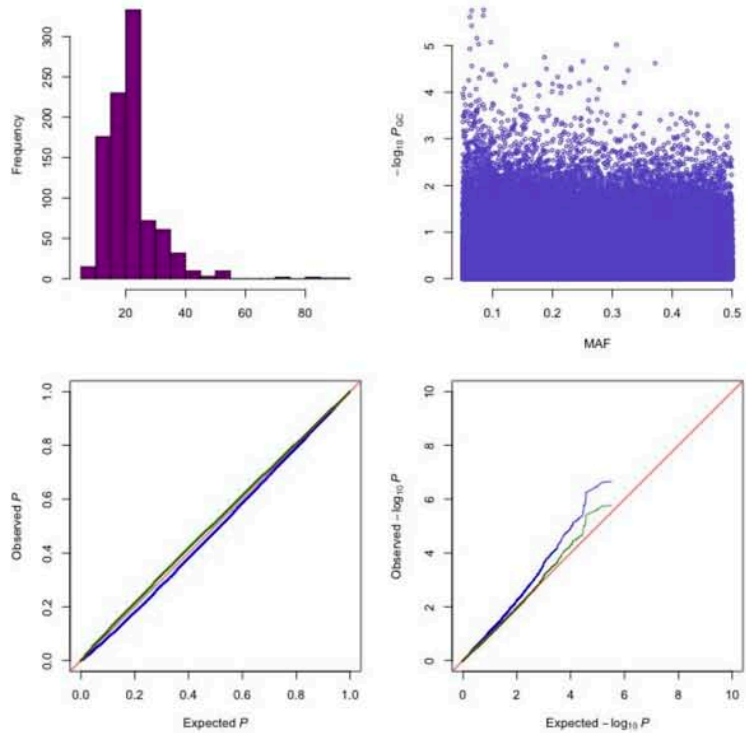

### b: Genome-wide association mapping for climate adaptability.

The plotted $-\log_{10}p$-values are genomic controlled. Markers with minor allele frequencies less than 5% are removed. Chromosomes are distinguished by colors. The Bonferroni-corrected significance threshold is marked by the horizontal line.

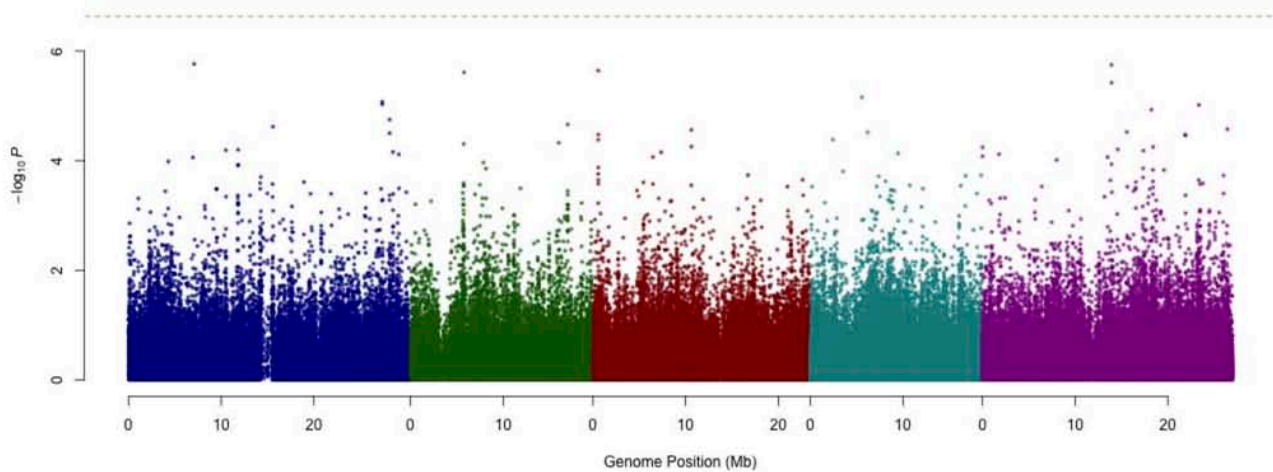

**Supplementary Figure 6** - Summary of results for precipitation CV.

### a: Phenotypic and *p*-value distributions.

Top-left: phenotypic distribution; Top-right: $-\log_{10}p$-values after genomic control (GC) against minor allele frequencies (MAF); Bottom panels: Quantile-quantile plots of *p*-values and $-\log_{10}p$-values before (blue) and after (green) GC.

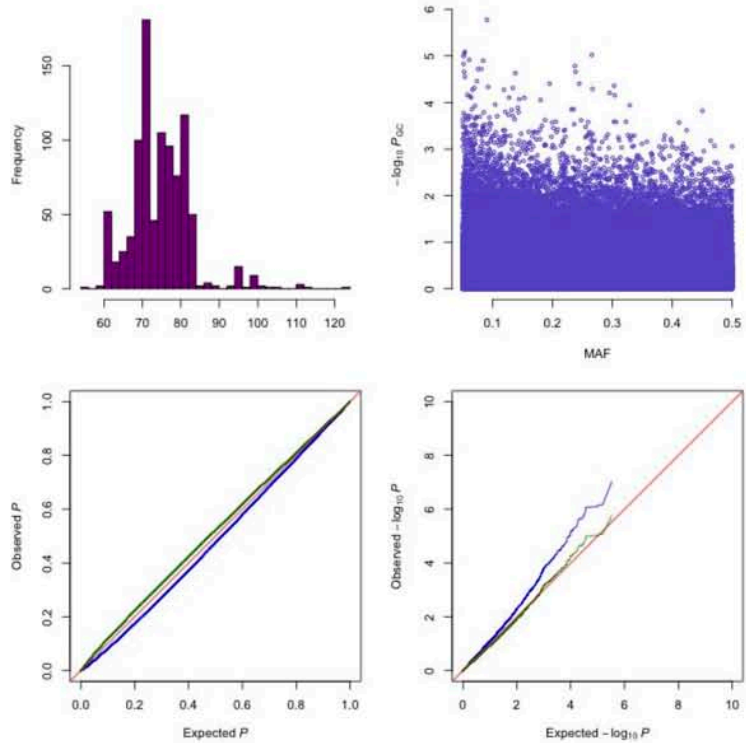

### b: Genome-wide association mapping for climate adaptability.

The plotted $-\log_{10}p$-values are genomic controlled. Markers with minor allele frequencies less than 5% are removed. Chromosomes are distinguished by colors. The Bonferroni-corrected significance threshold is marked by the horizontal line.

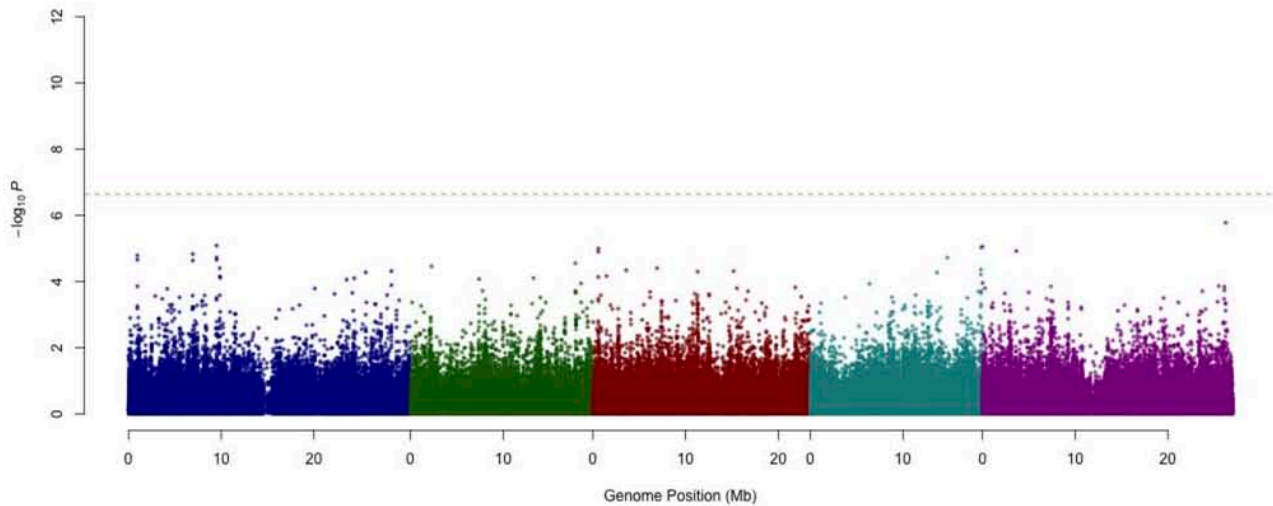

**Supplementary Figure 7** - Summary of results for photosynthetically active radiation in spring.

### a: Phenotypic and *p*-value distributions.

Top-left: phenotypic distribution; Top-right: $-\log_{10}p$-values after genomic control (GC) against minor allele frequencies (MAF); Bottom panels: Quantile-quantile plots of *p*-values and $-\log_{10}p$-values before (blue) and after (green) GC.

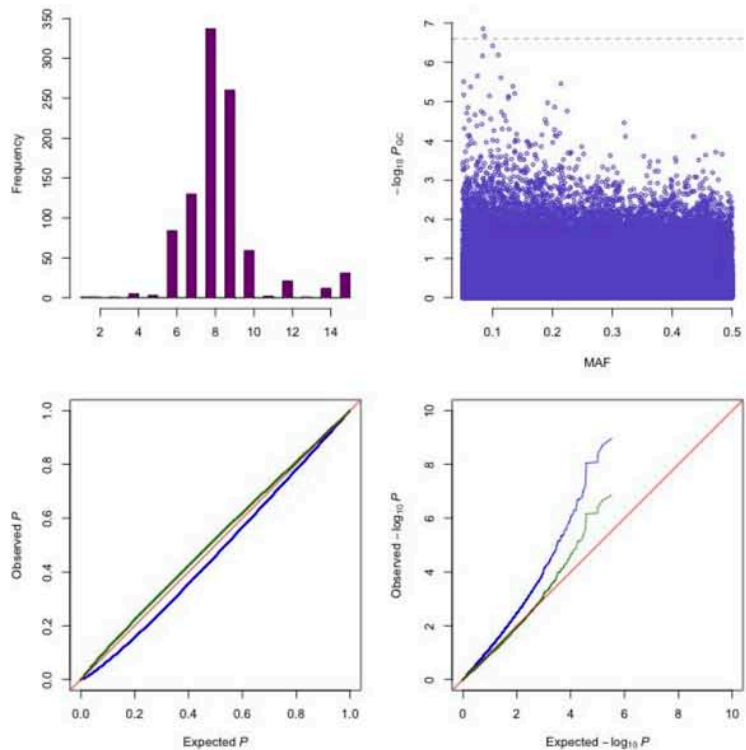

### b: Genome-wide association mapping for climate adaptability.

The plotted $-\log_{10}p$-values are genomic controlled. Markers with minor allele frequencies less than 5% are removed. Chromosomes are distinguished by colors. The Bonferroni-corrected significance threshold is marked by the horizontal line.

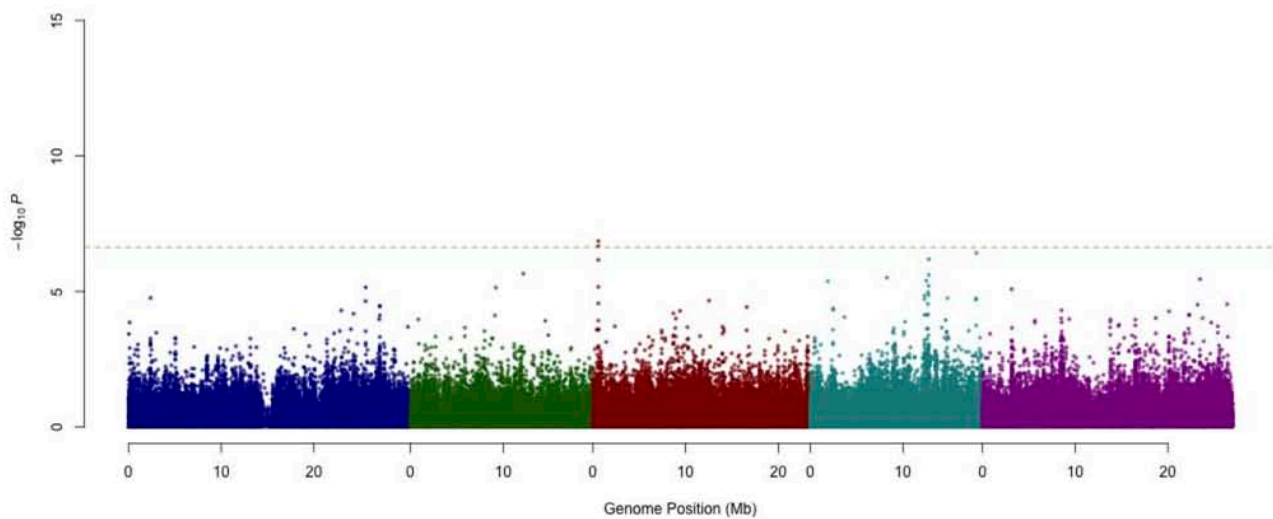

**Supplementary Figure 8** - Summary of results for length of the growing season.

**a: Phenotypic and *p*-value distributions.**

Top-left: phenotypic distribution; Top-right: $-\log_{10} p$-values after genomic control (GC) against minor allele frequencies (MAF); Bottom panels: Quantile-quantile plots of *p*-values and $-\log_{10} p$-values before (blue) and after (green) GC.

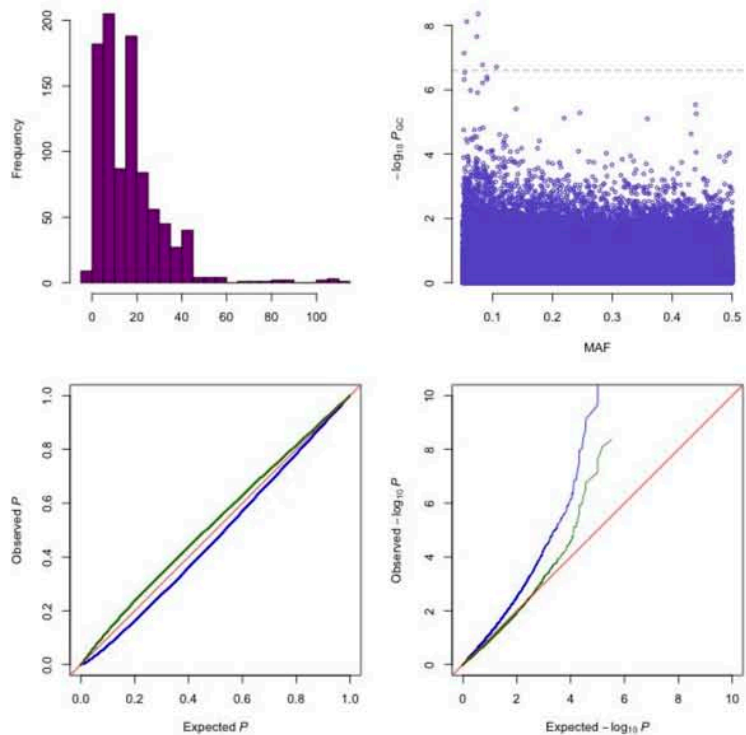

**b: Genome-wide association mapping for climate adaptability.**

The plotted $-\log_{10} p$-values are genomic controlled. Markers with minor allele frequencies less than 5% are removed. Chromosomes are distinguished by colors. The Bonferroni-corrected significance threshold is marked by the horizontal line.

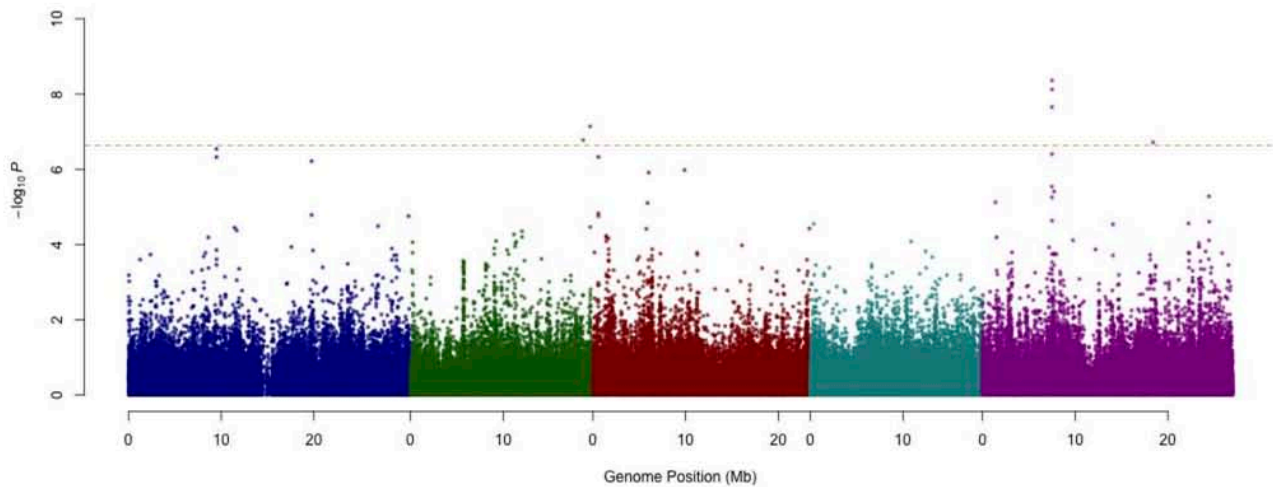

**Supplementary Figure 9** - Summary of results for number of consecutive cold days.

### a: Phenotypic and *p*-value distributions.

Top-left: phenotypic distribution; Top-right: -$\log_{10}p$-values after genomic control (GC) against minor allele frequencies (MAF); Bottom panels: Quantile-quantile plots of *p*-values and -$\log_{10}p$-values before (blue) and after (green) GC.

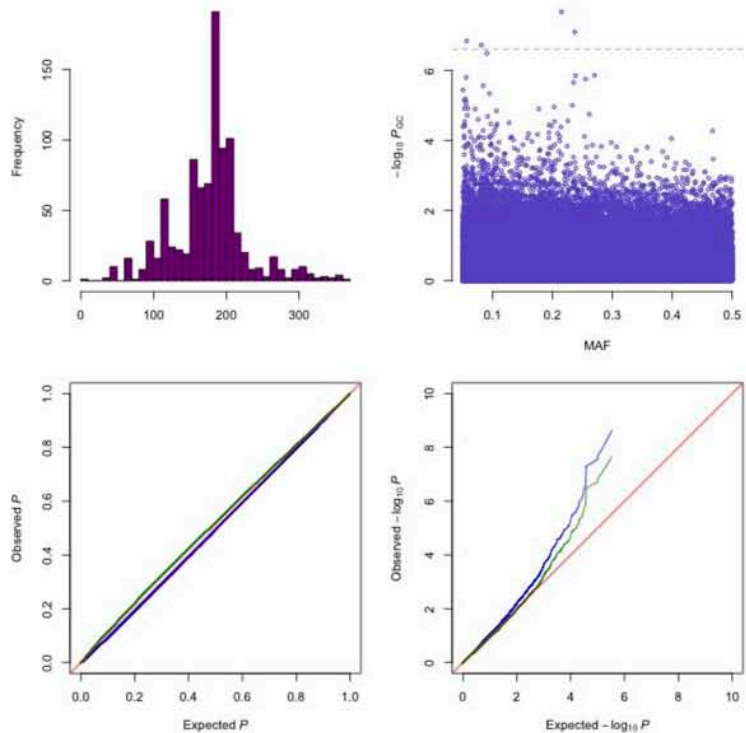

### b: Genome-wide association mapping for climate adaptability.

The plotted -$\log_{10}p$-values are genomic controlled. Markers with minor allele frequencies less than 5% are removed. Chromosomes are distinguished by colors. The Bonferroni-corrected significance threshold is marked by the horizontal line.

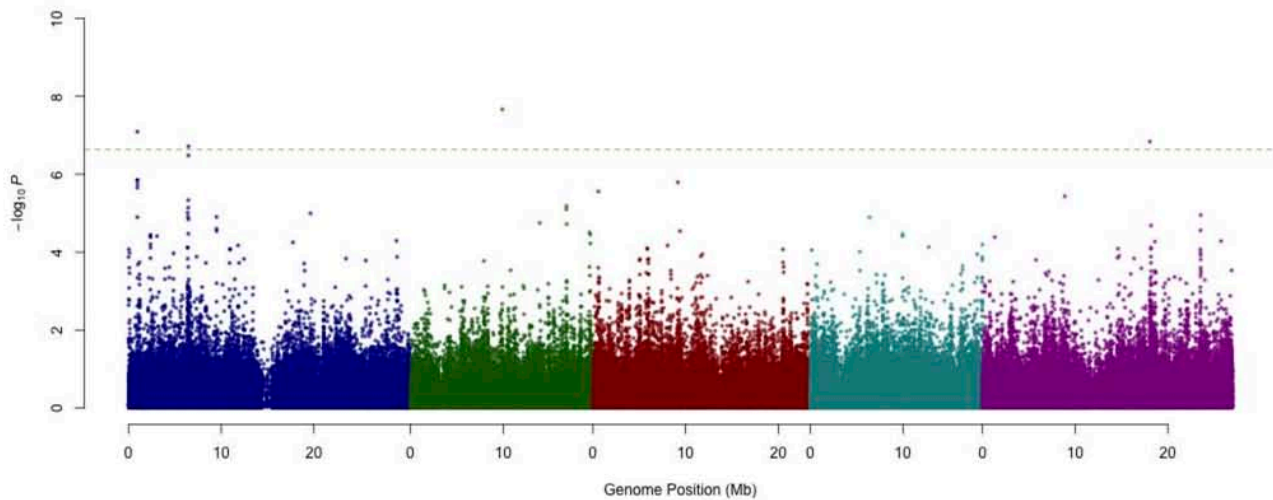

**Supplementary Figure 10** - Summary of results for number of consecutive frost-free days.

### a: Phenotypic and *p*-value distributions.

Top-left: phenotypic distribution; Top-right: -$\log_{10}p$-values after genomic control (GC) against minor allele frequencies (MAF); Bottom panels: Quantile-quantile plots of *p*-values and -$\log_{10}p$-values before (blue) and after (green) GC.

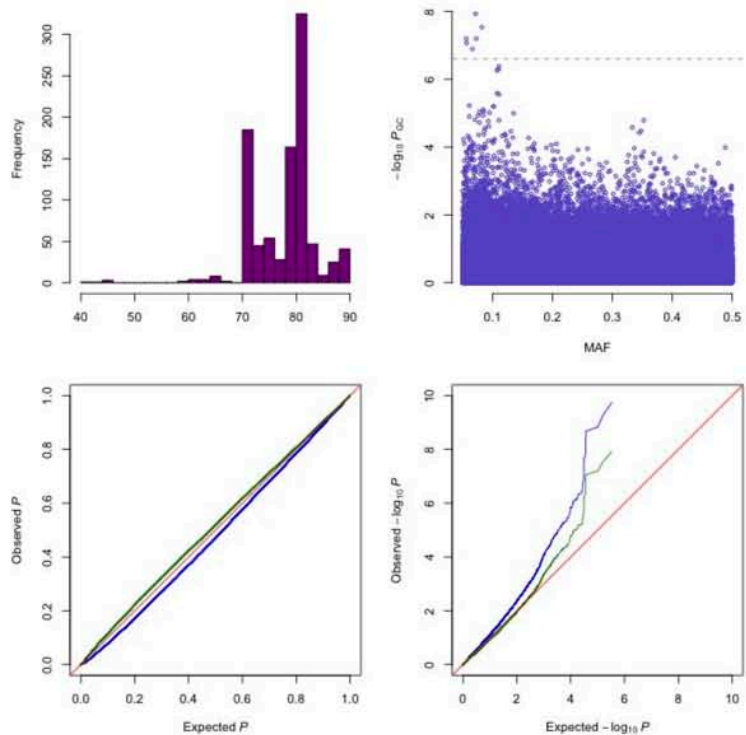

### b: Genome-wide association mapping for climate adaptability.

The plotted -$\log_{10}p$-values are genomic controlled. Markers with minor allele frequencies less than 5% are removed. Chromosomes are distinguished by colors. The Bonferroni-corrected significance threshold is marked by the horizontal line.

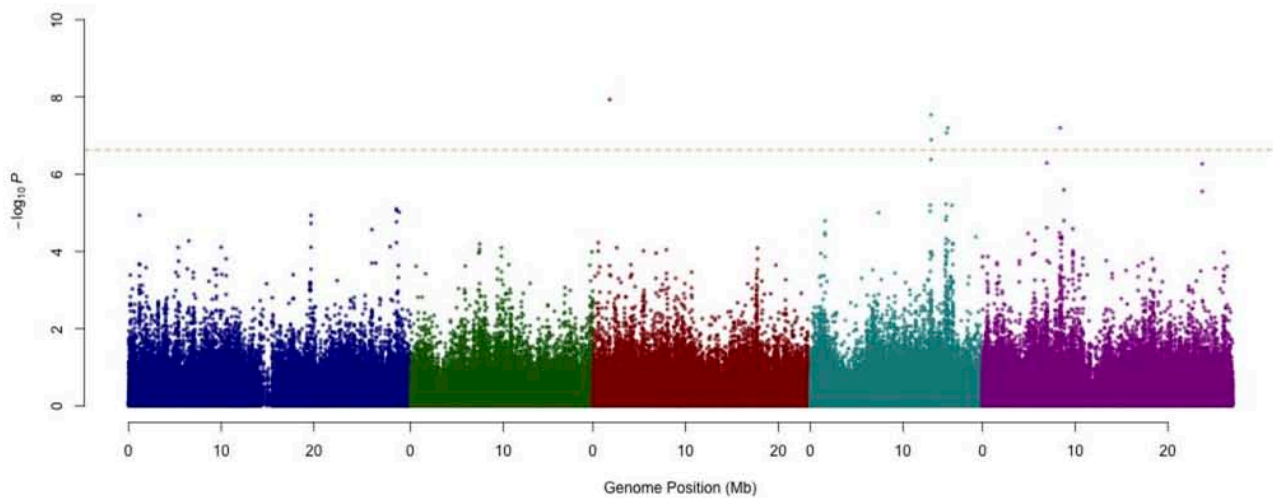

**Supplementary Figure 11** - Summary of results for relative humidity in spring.

### a: Phenotypic and *p*-value distributions.

Top-left: phenotypic distribution; Top-right: $-\log_{10}p$-values after genomic control (GC) against minor allele frequencies (MAF); Bottom panels: Quantile-quantile plots of *p*-values and $-\log_{10}p$-values before (blue) and after (green) GC.

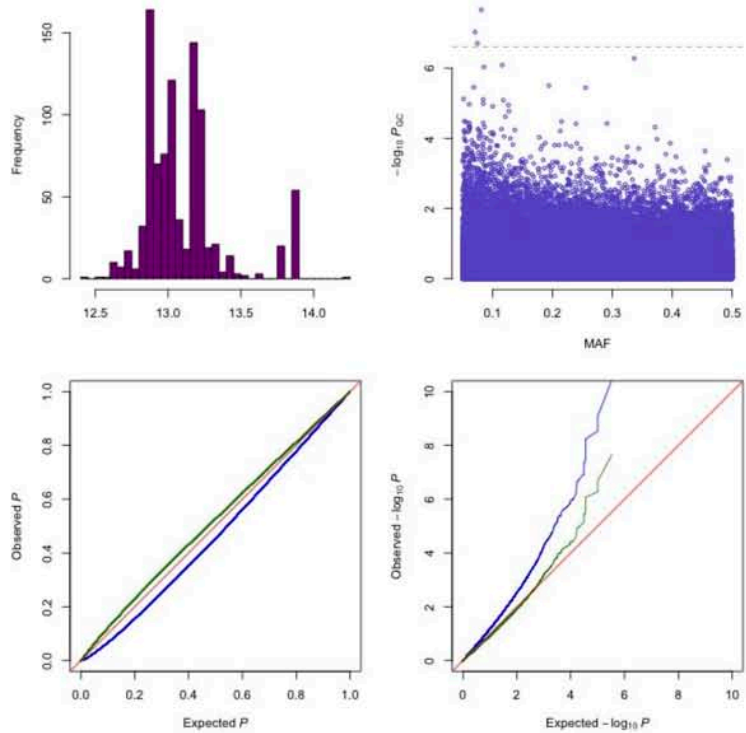

### b: Genome-wide association mapping for climate adaptability.

The plotted $-\log_{10}p$-values are genomic controlled. Markers with minor allele frequencies less than 5% are removed. Chromosomes are distinguished by colors. The Bonferroni-corrected significance threshold is marked by the horizontal line.

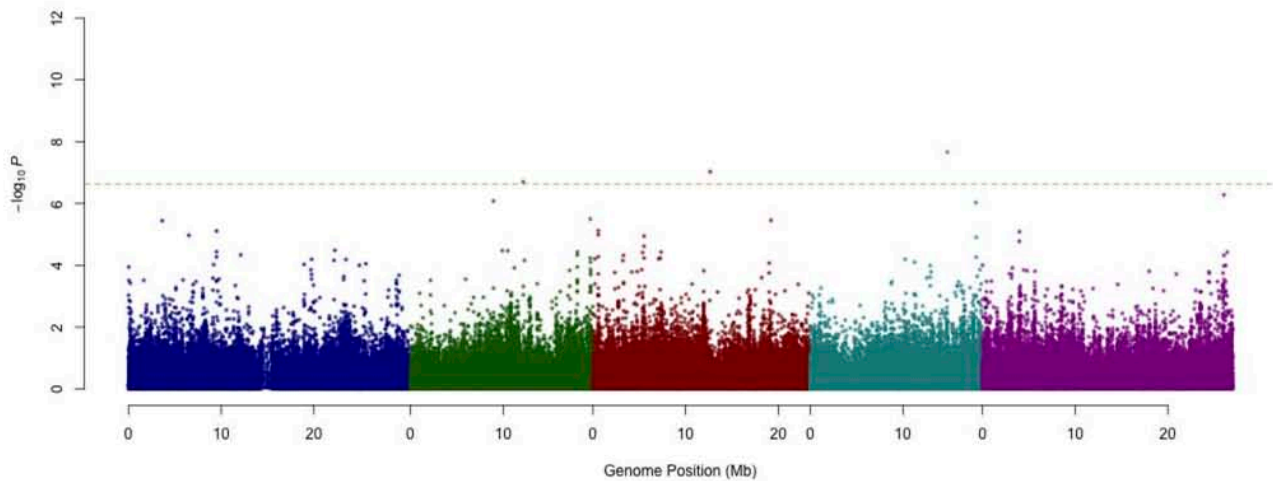

**Supplementary Figure 12** - Summary of results for daylength in spring.

## a: Phenotypic and *p*-value distributions.

Top-left: phenotypic distribution; Top-right: $-\log_{10}p$-values after genomic control (GC) against minor allele frequencies (MAF); Bottom panels: Quantile-quantile plots of *p*-values and $-\log_{10}p$-values before (blue) and after (green) GC.

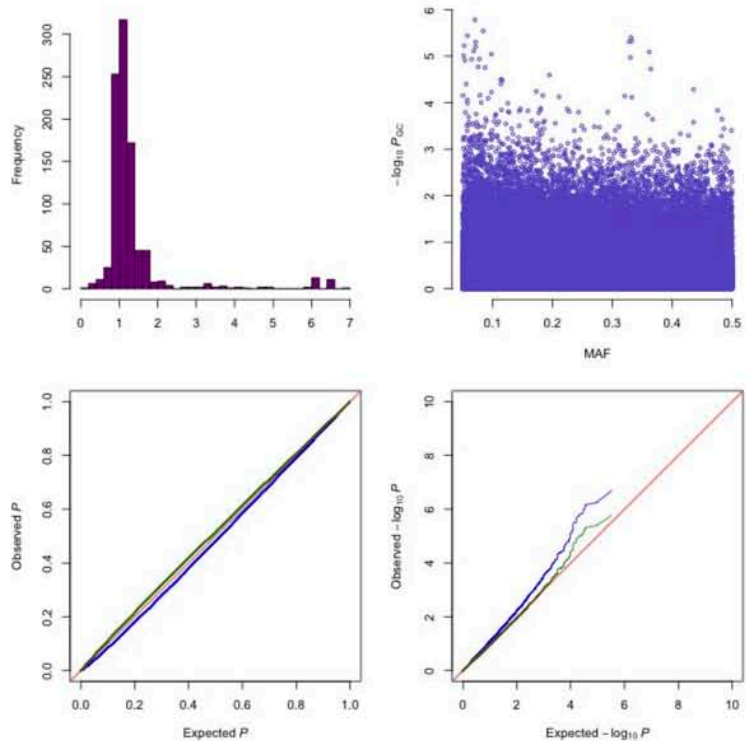

## b: Genome-wide association mapping for climate adaptability.

The plotted $-\log_{10}p$-values are genomic controlled. Markers with minor allele frequencies less than 5% are removed. Chromosomes are distinguished by colors. The Bonferroni-corrected significance threshold is marked by the horizontal line.

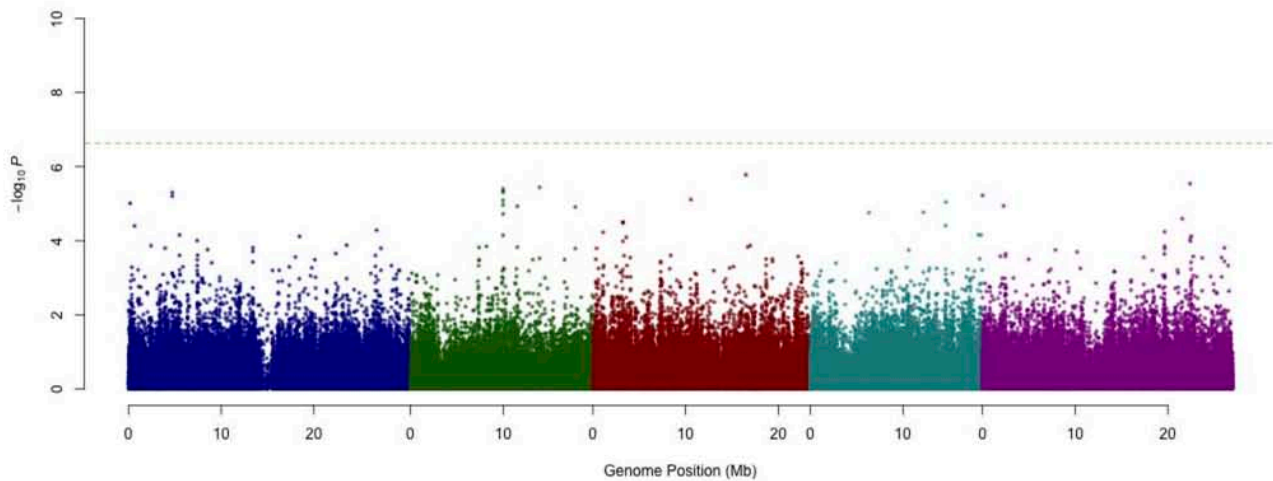

**Supplementary Figure 13** - Summary of results for aridity index.

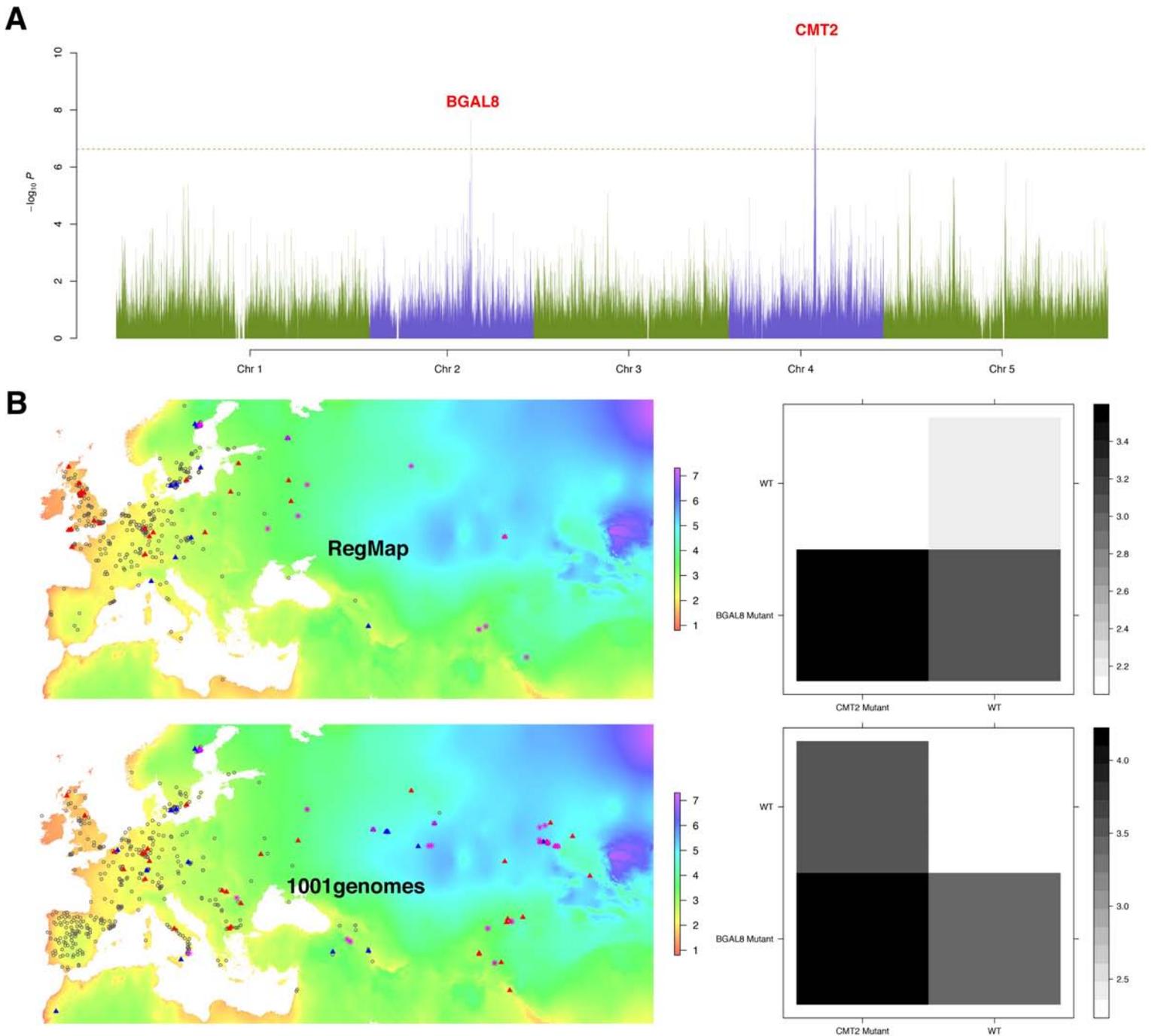

**Supplementary Figure 14: Epistatic interaction between the two significant loci CMT2 and BGAL8 for temperature seasonality.** (**A**) Both loci were significantly mapped in genome-wide association analysis for adaptability in the RegMap sample at a Bonferroni-corrected significant threshold (horizontal dashed line), corrected for population stratification. (**B**) The epistasis genotype-phenotype map was shown in the Euro-Asia map and a CMT2 × BGAL8 heatmap (colorkey represents temperature seasonality) for both the RegMap collection and the 1001-genomes accessions. WT: wild-type. In the Euro-Asia map, gray circles are the accessions being WT/WT at both loci; red triangles only carry the CMT2 mutation and blue only carry the BGAL8 mutation; the pink stars are accessions carrying both mutations.

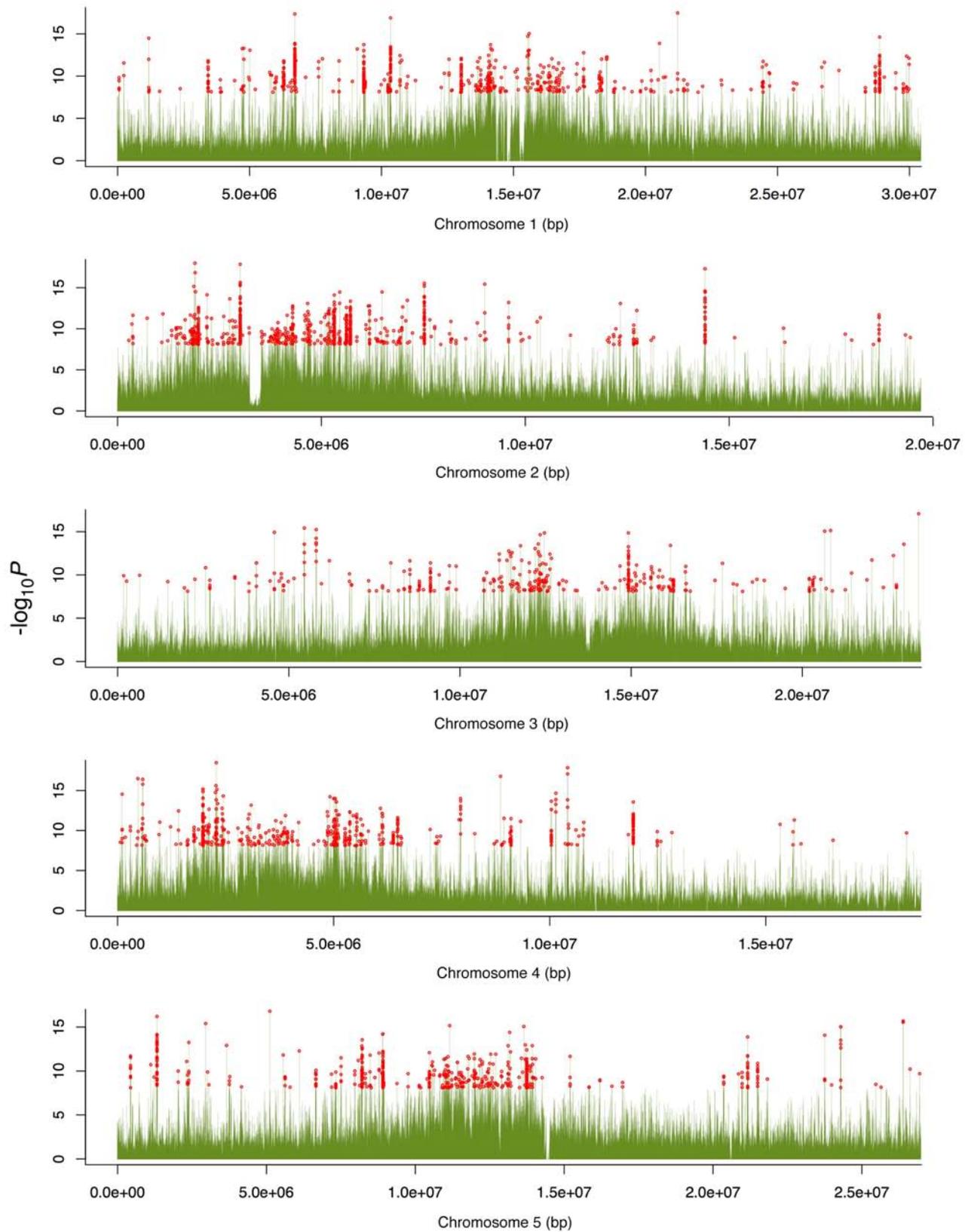

**Supplementary Figure 15: Methylome-wide association tests for CMT2$_{STOP}$ genotypes.** The significant methylation sites passing the Bonferroni-corrected significance threshold are marked in red, which were used in the validation analysis in the CMT2 knockouts.

**Supplementary table 1: Mis- and non-sense mutations in high-LD with genome-wide significant, non-additive associations to climate adaptability.**

| Trait | Chrom | Pos (bp) | Locus | Gene name | Consequence | Protein position | AA Change | Codon change | MAF | Mutant analysis PASE | MSA |
|---|---|---|---|---|---|---|---|---|---|---|---|
| Temperature seasonality | | | | | | | | | | | |
| | 4 | 10 405 599 | AT4G19000 | IWS2 | missense | 33 | S/T | Tcg/Acg | 0.11 | 0.21 | 0.35 |
| | 4 | 10 410 039 | AT4G19006 | | missense | 240 | H/L | cAc/cTc | 0.11 | 0.52 | 0.41 |
| | 4 | 10 411 801 | AT4G19010 | | missense | 539 | N/H | Aac/Cac | 0.11 | 0.29 | 0.08 |
| | 4 | 10 412 877 | AT4G19010 | | missense | 373 | S/L | tCg/tTg | 0.11 | 0.52 | 0.14 |
| | 4 | 10 413 290 | AT4G19010 | | missense | 311 | A/V | gCa/gTa | 0.12 | 0.35 | 0.68 |
| | 4 | 10 413 374 | AT4G19010 | | missense | 283 | K/R | aAg/aGg | 0.12 | 0.23 | 0.2 |
| | 4 | 10 414 556 | AT4G19020 | CMT2 | stop | 11 | E/* | Gag/Tag | 0.11 | STOP | STOP |
| | 4 | 10 414 640 | AT4G19020 | CMT2 | frameshift | 39 | - | - | 0.00 | FS | FS |
| | 4 | 10 414 640 | AT4G19020 | CMT2 | missense | 39 | E/K | Gaa/Aaa | 0.12 | 0.63 | 0.17 |
| | 4 | 10 414 747 | AT4G19020 | CMT2 | missense | 74 | N/K | aaC/aaG | 0.11 | 0.45 | 0.11 |
| | 4 | 10 415 805 | AT4G19020 | CMT2 | missense | 354 | G/D | gGc/gAc | 0.11 | 0.58 | 0.11 |
| | 4 | 10 415 833 | AT4G19020 | CMT2 | missense | 363 | L/F | ttA/ttT | 0.11 | 0.2 | 0.14 |
| | 4 | 10 416 047 | AT4G19020 | CMT2 | missense | 435 | G/S | Ggt/Agt | 0.11 | 0.26 | 0.14 |
| | 4 | 10 445 425 | AT4G19060 | | missense | 270 | I/T | aTt/aCt | 0.11 | 0.37 | 0.44 |
| | 4 | 10 447 160 | AT4G19070 | | missense | 140 | S/R | Agt/Cgt | 0.11 | 0.73 | 0.11 |
| | 4 | 10 450 001 | AT4G19090 | | missense | 17 | R/G | Aga/Gga | 0.11 | 0.93 | 0.07 |
| | 4 | 10 455 784 | AT4G19110 | | missense | 192 | L/F | ttG/ttT | 0.10 | 0.2 | 0.45 |
| | 4 | 10 455 784 | AT4G19110 | | missense | 248 | L/F | ttG/ttT | 0.10 | 0.2 | 0.44 |
| | 4 | 10 455 784 | AT4G19110 | | missense | 248 | L/F | ttG/ttT | 0.10 | 0.2 | 0.44 |
| Maximum temperature in the warmest month | | | | | | | | | | | |
| | 1 | 6 936 457 | AT1G19990 | | missense | 183 | S/F | tCt/tTt | 0.09 | 0.64 | 0.2 |
| Minimum temperature in the coldest month | | | | | | | | | | | |
| | 5 | 14 067 526 | AT5G35930 | | missense | 963 | E/Q | Gaa/Caa | 0.12 | 0.3 | 0.05 |
| Number of consecutive cold days | | | | | | | | | | | |
| | 5 | 7 492 033 | AT5G22560 | | missense | 355 | E/D | gaG/gaC | 0.14 | 0.26 | 0.22 |
| | 5 | 7 492 259 | AT5G22560 | | missense | 280 | R/T | aGa/aCa | 0.13 | 0.63 | 0.11 |
| | 5 | 7 492 277 | AT5G22560 | | missense | 274 | N/T | aAt/aCt | 0.14 | 0.27 | 0.11 |
| | 5 | 7 492 277 | AT5G22560 | | missense | 274 | N/S | aAt/aGt | 0.14 | 0.3 | 0.11 |
| Relative humidity in spring & Day length in spring | | | | | | | | | | | |
| | 4 | 14 788 320 | AT4G30200 | VEL1 | missense | 236 | E/D | gaA/gaT | 0.11 | 0.26 | 0.06 |
| | 4 | 14 788 320 | AT4G30200 | VEL1 | missense | 253 | E/D | gaA/gaT | 0.11 | 0.26 | 0.06 |
| | 4 | 14 829 581 | AT4G30290 | XTH19 | missense | 100 | I/V | Att/Gtt | 0.07 | 0.14 | 0.49 |
| Minimum temperature in the coldest month & Number of consecutive cold days | | | | | | | | | | | |
| | 2 | 19 397 389 | AT2G47240 | | missense | 620 | K/R | aAa/aGa | 0.06 | 0.23 | 0.13 |
| | 2 | 19 397 389 | AT2G47240 | | missense | 620 | K/R | aAa/aGa | 0.06 | 0.23 | 0.13 |
| Temperature seasonality & Day length in spring | | | | | | | | | | | |
| | 2 | 12 169 734 | AT2G28470 | BGAL8 | missense | 672 | E/D | gaA/gaT | 0.11 | 0.26 | 0.06 |
| | 2 | 12 169 734 | AT2G28470 | BGAL8 | missense | 678 | E/D | gaA/gaT | 0.11 | 0.26 | 0.06 |
| | 2 | 12 169 828 | AT2G28470 | BGAL8 | missense | 647 | F/Y | tTc/tAc | 0.12 | 0.36 | 0.72 |
| | 2 | 12 169 828 | AT2G28470 | BGAL8 | missense | 641 | F/Y | tTc/tAc | 0.12 | 0.36 | 0.72 |
| Number of consecutive frost-free days | | | | | | | | | | | |
| | 1 | 954 782 | AT1G03790 | SOM | missense | 65 | N/I | aAt/aTt | 0.25 | 0.6 | 0.06 |
| | 1 | 955 189 | AT1G03790 | SOM | missense | 201 | P/T | Cct/Act | 0.25 | 0.22 | 0.17 |
| | 1 | 955 268 | AT1G03790 | SOM | missense | 227 | S/C | tCt/tGt | 0.25 | 0.55 | 0.4 |

**Supplementary table 2: Genes located less than 100Kb up- or down-stream of the leading SNP in the Genome-Wide Association analysis and that also are in high linkage disequilibrium with the SNP ($r^2 > 0.8$)**

| Trait | Gene |
|---|---|
| Temperature seasonality | |
| | AT4G18960 AG K-box region and MADS-box transcription factor family protein |
| | AT4G18970 GDSL-like Lipase/Acylhydrolase superfamily protein |
| | AT4G18975 Pentatricopeptide repeat (PPR) superfamily protein |
| | AT4G18980 AtS40-3 |
| | AT4G18990 XTH29 xyloglucan endotransglucosylase/hydrolase 29 |
| | AT4G19003 VPS25 E2F/DP family winged-helix DNA-binding domain |
| | AT4G19030 NLM1 NOD26-like major intrinsic protein 1 |
| | AT4G19035 LCR7 low-molecular-weight cysteine-rich 7 |
| | AT4G19038 LCR15 low-molecular-weight cysteine-rich 15 |
| | AT4G19040 EDR2 ENHANCED DISEASE RESISTANCE 2 |
| | AT4G19045 Mob1/phocein family protein |
| | AT4G19050 NB-ARC domain-containing disease resistance protein |
| | AT4G19080 unknown protein |
| | AT4G19095 unknown protein |
| | AT4G19100 unknown protein |
| | AT4G19112 CPuORF25 conserved peptide upstream open reading frame 25 |
| | AT4G19120 ERD3 S-adenosyl-L-methionine-dependent methyltransferases superfamily protein |
| Maximum temperature in the warmest month | |
| | AT1G19970 ER lumen protein retaining receptor family protein |
| | AT1G19980 cytomatrix protein-related |
| | AT1G20000 TAF11b TBP-associated factor 11B |
| | AT1G20010 TUB5 tubulin beta-5 chain |
| | AT1G20015 snoRNA |
| Minimum temperature in the coldest month | |
| | AT5G35926 Protein with RNI-like/FBD-like domains |
| Number of consecutive cold days | |
| | AT5G22555 unknown protein |
| | AT5G22570 WRKY38 WRKY DNA-binding protein 38 |
| Day length in spring | |
| | AT3G30859 transposable element gene |
| | AT3G30867 pseudogene, putative SNF8 protein homolog |
| Relative humidity in spring & Day length in spring | |
| | AT4G30240 Syntaxin/t-SNARE family protein |
| | AT4G30250 P-loop containing nucleoside triphosphate hydrolases superfamily protein |
| | AT4G30260 Integral membrane Yip1 family protein |
| | AT4G30270 MERI5B xyloglucan endotransglucosylase/hydrolase 24 |
| | AT4G30280 ATXTH18 xyloglucan endotransglucosylase/hydrolase 18 |
| | AT4G30300 ATNAP15 non-intrinsic ABC protein 15 |
| | AT4G30320 CAP (Cysteine-rich secretory proteins, Antigen 5, and Pathogenesis-related 1 protein) superfamily protein |
| | AT4G30330 Small nuclear ribonucleoprotein family protein |
| | AT4G30340 ATDGK7 diacylglycerol kinase 7 |
| Minimum temperature in the coldest month & Number of consecutive cold days | |
| | AT2G47250 RNA helicase family protein |
| | AT2G45150 CDS4 cytidinediphosphate diacylglycerol synthase 4 |
| | AT2G45160 HAM1 GRAS family transcription factor |
| | AT2G45161 unknown protein |
| | AT2G45170 ATATG8E AUTOPHAGY 8E |
| | AT5G45380 ATDUR3 solute:sodium symporters;urea transmembrane transporters |
| | AT5G45390 CLPP4 CLP protease P4 |
| | AT5G45400 RPA70C Replication factor-A protein 1-related |
| | AT5G45410 unknown protein |
| Temperature seasonality & Day length in spring | |
| | AT2G28410 unknown protein |
| | AT2G28420 Lactoylglutathione lyase / glyoxalase I family protein |
| | AT2G28426 unknown protein |
| | AT2G28430 unknown protein |
| | AT2G28440 proline-rich family protein |
| | AT2G28450 zinc finger (CCCH-type) family protein |
| | AT2G28460 Cysteine/Histidine-rich C1 domain family protein |
| Relative humidity in spring | |
| | AT3G06019 unknown protein |
| | AT3G06020 unknown protein |
| | AT3G06030 ANP3 NPK1-related protein kinase 3 |
| | AT5G24530 DMR6 2-oxoglutarate (2OG) and Fe(II)-dependent oxygenase superfamily protein |
| | AT5G24540 BGLU31 beta glucosidase 31 |
| Length of the growing season | |
| | AT3G02660 Tyrosyl-tRNA synthetase, class Ib, bacterial/mitochondrial |
| | AT3G02670 Glycine-rich protein family |
| | AT3G02680 NBS1 nijmegen breakage syndrome 1 |
| | AT3G02690 nodulin MtN21 /EamA-like transporter family protein |
| Number of consecutive frost-free days | |
| | AT1G03780 TPX2 targeting protein for XKLP2 |
| | AT1G03800 ERF10 ERF domain protein 10 |
| | AT1G03810 Nucleic acid-binding, OB-fold-like protein |
| | AT1G03820 unknown protein |
| | AT1G03830 guanylate-binding family protein |
| | AT1G18720 unknown protein |
| | AT1G18730 NDF6 NDH dependent flow 6 |
| | AT1G18735 other RNA |
| | AT1G18740 unknown protein |
| | AT1G18745 NcRNA |
| | AT1G18750 AGL65 AGAMOUS-like 65 |
| | AT2G23250 UGT84B2 UDP-glucosyl transferase 84B2 |
| | AT2G23260 UGT84B1 UDP-glucosyl transferase 84B1 |
| | AT2G23270 unknown protein |
| | AT2G23290 AtMYB70 myb domain protein 70 |